\documentclass[format=acmsmall,nonacm=true]{acmart} 

\usepackage{cleveref} 
\Crefname{section}{Section}{Sections}
\Crefname{table}{Table}{Tables}
\Crefname{figure}{Figure}{Figures}
\usepackage{booktabs}
\usepackage{multirow}
\usepackage{tabularx}
\usepackage{ragged2e}  %
\newcolumntype{L}[1]{>{\hsize=#1\hsize\raggedright\arraybackslash}X}
\newcolumntype{P}[1]{>{\raggedright\arraybackslash}p{#1}}
\newcolumntype{?}{!{\vrule width 1pt}}
\usepackage{paralist}
\usepackage{subfig}

\newcommand{\wrt}{with respect to\ }
\newcommand{\cf}{cf.\ } 		
\newcommand{\ies}{i.e.,\ } 		
\newcommand{\ie}{that is,\ }			
\newcommand{\egs}{e.g.\ } 		
\newcommand{\Eg}{For example,\ }
\newcommand{\eg}{for example,\ }

\newcommand{\rb}[2][90]{\rotatebox{#1}{\parbox{2.3cm}{\emph{#2}}}}

\makeatletter
\newcommand{\trackedcontentNamed}[3]{%
  \emph{#3}%
  \def\@currentlabel{#3}%
  \label{#1#2}%
}
\newcommand{\RQ}[2]{q\trackedcontentNamed{SQ-}{#1}{#2}}
\makeatother
\newcommand{\RQRef}[1]{q\ref{SQ-#1}}

\newcommand{\trackedcontent}[4]{%
  \refstepcounter{#1}%
  #4\arabic{#1}: #3%
  \label{#2}}
\newcommand{\Metacite}[1]{%
  \trackedcontent{tcMetacite}{cit:#1}{\cite{#1}}{}}
\newcommand{\MetaciteRef}[1]{\ref{cit:#1}}

\usepackage{tikz}
\usetikzlibrary{arrows,arrows.meta,positioning,calc}

\setcitestyle{sort&compress}

\acmArticle{}

\begin{document}

\title[Assurance of System Safety: Designs and Arguments]
{Assurance of System Safety:\newline A Survey of Design and Argument Patterns}
\author{Mario Gleirscher} %
\authornote{Corresponding author.}
\orcid{0000-0002-9445-6863}
\affiliation{%
  \institution{University of York}
  \department{Department of Computer Science}
  \streetaddress{Deramore Lane}
  \city{Heslington, York}
  \state{Yorkshire}
  \postcode{YO10 5GH}
  \country{United Kingdom}}
\email{mario.gleirscher@york.ac.uk}
\author{Stefan Kugele}
\affiliation{%
  \institution{Technical University of Munich}
  \department{Department of Informatics}
  \streetaddress{Boltzmannstr. 3}
  \city{Garching bei M\"unchen}
  \state{Bavaria}
  \postcode{85748}
  \country{Germany}}
\email{stefan.kugele@tum.de}

\thanks{Supported by the
  \grantsponsor{501100001659}{Deutsche Forschungsgemeinschaft
    (DFG)}{http://doi.org/10.13039/501100001659} under 
  Grants~no.~\grantnum{501100001659}{GL~915/1-1} and~\grantnum{501100001659}{GL~915/1-2}.
  \copyright\ 2018. This manuscript is
  made available under the CC-BY-NC-ND 4.0 license
  \url{http://creativecommons.org/licenses/by-nc-nd/4.0/}.
  \newline
  \textbf{Reference Format:}
  Gleirscher, M., \& Kugele, S.. \emph{Assurance of System Safety: A Survey of Design
    and Argument Patterns} (\today). Unpublished working paper. 
  Department of Computer Science, University of York, United
  Kingdom.
  \href{https://arxiv.org/abs/?}{arXiv:id [cs.SE]}
}

\begin{abstract}
  The specification, design, and assurance of safety encompasses
  various concepts and best practices, subject of reuse in
  form of \emph{patterns}.
  This work summarizes applied research on such concepts and practices
  with a focus on the last two decades and on the state-of-the-art of
  patterns in safety-critical system design and assurance
  argumentation.
  We investigate several aspects of such patterns, \eg where and when
  they are applied, their characteristics and purposes, and how they
  are related.
  For each aspect, we provide an overview of relevant studies and
  synthesize a taxonomy of first principles underlying these patterns.
  Furthermore, we comment on how these studies address known
  challenges and we discuss suggestions for further research.
  Our findings disclose a lack of research on how patterns improve
  system safety claims and, vice versa, on the decomposition of system
  safety into separated local concerns, and on the impact of security
  on safety.
\end{abstract}



\maketitle

\renewcommand{\shortauthors}{M.\ Gleirscher et al.}

\section{Introduction}
\label{sec:introduction}

Safety is an indisputably critical and strongly desirable property of
engineered systems as operated in their designated
environments~\citep{McDermid1991,Leveson2012}.  The assurance of this
property remains a critical activity throughout the life cycle of such
systems.  The procedure and the key elements of safety engineering are
reflected in generic and domain-specific methods, techniques, and
standards.

\subsection{Background}
\label{sec:background}

This section highlights important concepts in safety assurance,
introduces the terminology used below, and summarizes core aspects of
design and argument patterns used in safety assurance.

\subsubsection{Systems and Safety}
\label{sec:viewsofsafety}
\label{sec:safety}

Consider an engineered physical system %
operated in a domestic, urban, or industrial environment, \eg an
autonomous mobile robot carrying through tasks in a warehouse---for
the sake of discussing a recent application domain.

By \emph{system safety}~(safety for short), we refer to the extent to
which such a system is free of hazards, that is, of risks of physical
harm for humans, the environment, the system itself, and other usually
physical
assets~\citep{Burns1992-MeaningSafetySecurity,Leveson2012,Lund2011}.
In the robot example, such hazards could involve collisions of the
robot with obstacles, the robot falling down the stairs, the robot
dumping down valuables, or the robot having an internal \emph{component
failure} or getting malicious or erroneous control inputs potentially
triggering one of the previous risks.

One way of classifying failures is to distinguish systematic from
random causes.  Systematic faults are associated with development
mistakes leading to wrong specifications, designs, or implementations.
Materials such as, \eg mechanical,
electrical, and electronic components exhibit random
faults because of aging, degradation, or electromagnetic radiation.
According to a widespread view, random behavior of software can only
result from random inputs or faults in the electronic hardware the
software is
running on.  We will use the term \emph{fault} in this study, although
many of the discussions apply to the terms \emph{error} and
\emph{failure}, \ie undesired higher-level or downstream events caused
by faults~\cite{Laprie1992,Avizienis2004}.

Overall, system safety is about handling critical events and
their possible undesired consequences.  Risk analysis as one step
towards handling such events deals with the estimation of \emph{risk
  levels} (also: risk priority) of these events, \ie abstractions of
the expected loss or cost resulting from these events.
Domain-specific risk classifiers (also: risk matrices) help safety
analysts to rank risks usually by estimating and combining two
parameters, \emph{probability of occurrence} and \emph{severity of
  consequence}.

In regulations (\egs in the United Kingdom and in
Australia), an engineered system is considered \emph{safe} if all
relevant risks have been reduced to a tolerable level, \ie a level at
which the cost of further risk reduction measures would be grossly
disproportionate to their benefit.

\subsubsection{Engineering Steps}
\label{sec:engineeringsteps}

The robot example indicates the many dimensions of system safety and
its tight relationship to other disciplines, \eg IT security.
Safety assurance considers the whole \emph{life cycle} of a system
(\ies the system as specified, designed, implemented, operated, and
decommissioned) as the primary source, root cause, or amplifier of
such risk and as the subject of assurance.  The \emph{system as
  operated} is the most tangible assurance object.

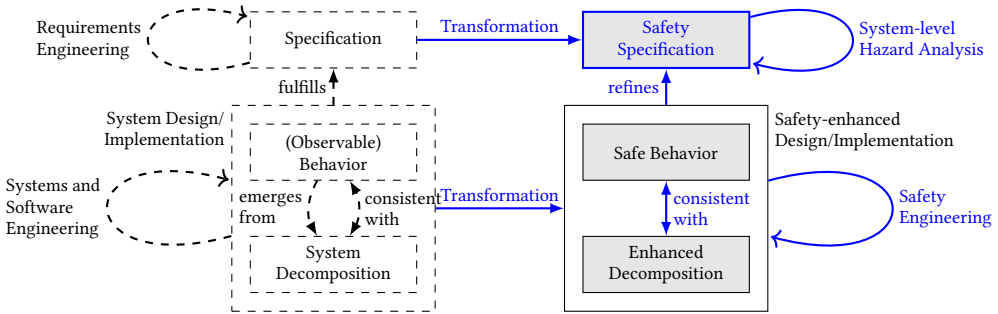
\begin{figure}[b]
  \centering
  \scriptsize
  \begin{tikzpicture}
    [dim/.style={minimum height=11em,minimum width=11em,draw,align=flush left},
    tit/.style={align=left,anchor=north west},
    itm/.style={draw,align=center,minimum height=3em,minimum
      width=9em},
    lab/.style={align=left},
    proc/.style={blue,thick,align=center},
    arg/.style={blue,thick,align=center}]

    \node[itm,dashed] (spec) at (0,0) {Specification};
    \node[dim,dashed] (dsgn) at ($(spec)+(0em,-9em)$) {};
    \node[itm,dashed] (beha) at ($(spec)+(0em,-6em)$) {(Observable)\\Behavior};
    \node[itm,dashed] (strc) at ($(beha)+(0em,-6em)$) {System\\Decomposition};
    \node[itm,arg,fill=gray!20] (spec-new) at ($(spec)+(18em,0)$) {Safety\\Specification};
    \node[dim] (dsgn-new) at ($(spec-new)+(0em,-9em)$) {};
    \node[itm,fill=gray!20] (beha-new) at ($(spec-new)+(0em,-6em)$) {Safe
      Behavior};
    \node[itm,fill=gray!20] (strc-new) at ($(beha-new)+(0em,-6em)$)
    {Enhanced\\Decomposition};

    \node[tit,align=right,anchor=north east] at (dsgn.north west) {System
      Design/\\Implementation};
    \node[tit] at (dsgn-new.north east) {Safety-enhanced\\Design/Implementation};

    \draw[arrows={-latex},thick,tips=proper]
    (spec) edge[dashed,loop left] node[lab] {Requirements\\Engineering} (spec)
    (dsgn) edge[dashed,loop left] node[lab] {Systems and \\Software\\Engineering} (dsgn)
    (spec-new) edge[proc,loop right] node[lab] {System-level\\Hazard Analysis} (spec-new)
    (dsgn-new) edge[proc,loop right] node[lab] {Safety\\Engineering} (dsgn-new)
    (beha) edge[dashed,left,bend right] node[lab] {emerges\\from} (strc) %
    (spec) edge[arg,above] node[lab] {Transformation} (spec-new)
    (dsgn) edge[arg,above] node[lab] {Transformation} (dsgn-new)
    (dsgn-new) edge[arg,left] node[lab] {refines} (spec-new)
    (dsgn) edge[dashed,left] node[lab] {fulfills} (spec)
    ;
    \draw[arrows={latex-latex},thick,tips=proper]
    (strc) edge[dashed,right,bend right] node[lab] {consistent\\with} (beha)
    (strc-new) edge[arg,right] node[lab] {consistent\\with} (beha-new)
    ;
  \end{tikzpicture}
  \caption{Assurance of system safety: core artifacts and their relationships
    \label{fig:overview}
  }
\end{figure}

In a typical life cycle, we distinguish the engineering steps of
specification, design, implementation, and assurance.  The left part
of \Cref{fig:overview} depicts the artifacts crafted in these steps.
For \textbf{specification}, the artifact ``specification'' is crafted
by the activity ``requirements engineering.''  For \textbf{design},
the artifact ``system design'' is crafted by the corresponding
``software and systems engineering'' activity.  We distinguish two
main abstractions used to model designs: \emph{behavior} and
\emph{decomposition}.
\textbf{Implementation} can be seen as a revision and refinement of
what is already there: the design.
The right part of \Cref{fig:overview} depicts the facets of
\textbf{assurance}~(blue arcs).  These facets usually comprise
\begin{itemize}
\item meeting the safety specification by guaranteeing reachability
  and invariance (the \emph{refines}-arc),
\item fault-avoidance and fault-tolerance (the
  \emph{transformation}-arc to ``enhanced decomposition,'' ``safe
  behavior,'' and ``safety specification,'' driven by system-level
  hazard analysis),
\item guidance from standards (all blue arcs),
\item case-based argumentation from verification, validation, and
  testing (all blue arcs).
\end{itemize}
\Cref{fig:overview} describes assurance as the argument (blue
arcs) that a consistently safety-enhanced implementation is a refinement
of a consistently safety-enhanced design which itself is a refinement
of a consistent safety specification.

\subsubsection{Patterns, Pattern Categories, Abstraction, and Tactics}
\label{sec:safetypatterns}

Models used in engineering---such as specifications, architectures, or
designs---are, if they address a common problem, candidates for
reuse~\citep{Kramer2007-IsAbstractionKey}.  We refer to
such abstractions as \emph{patterns}.
In this work, \emph{design and argument patterns} denote general
solutions for common problems recurring in the construction and
assurance practice of safety-critical systems.

The effective reusability of a pattern depends on the quality of its
\emph{documentation}.  In civil architecture~\citep{Alexander1977} and
later in software engineering~\citep{Gamma1993}, researchers started
to extract patterns from various sources and to catalog them using
\emph{templates} and \emph{models}.  Templates help in collecting
general information from the recurrences of a pattern such as, \eg the
context of use, the problem addressed, the solution, the consequences
of use, given names, underlying principles.  Models are useful to
capture technical details with a more expressive \emph{language}~(\egs
temporal logic, labeled transition system) and a standard
\emph{notation}~(\egs the Unified Modeling Language or the Goal
Structuring Notation).\footnote{See \url{http://www.omg.org}.}

Based on the artifacts and relationships in \Cref{fig:overview}, we
distinguish the following categories of patterns:

\begin{description}
\item[Specifications] %
  reflect practices to specify (\egs using domain-specific
  language \cite{Dwyer1999}) and decompose (\egs
  using contracts \cite{Meyer1992}) safety
  requirements.  Contracts
  form a practical way of including assumptions about the 
  environment of a system into a specification.
  \Eg\citet{Lamsweerde2009} provides a framework to
  construct specifications based on this idea.

\item[Designs] comprise two interrelated abstractions: (interface) behavior and
  (architectural) decomposition.
  
  \begin{description}
  \item[Behaviors] %
    describe safety concepts expressed in terms of a \emph{behavior}
    model, \eg a state machine encoding how a specific system is
    controlled to efficiently leave a dangerous situation or to enter
    a safe state.  If such behaviors are expressed \wrt the overall
    system---sharing the interface with the assets to be protected in
    the environment---then a corresponding contract can be
    formulated.

  \item[Decompositions] %
    represent reusable design practices to increase safety.  Such
    practices range from cross-disciplinary or mechatronic
    \emph{architecture} design to implementation in software,
    electronic, and mechanical hardware.  Common across these
    technologies is a particular \emph{decomposition} to implement
    principles such as, \eg monitoring, fault detection, redundancy,
    or recovery~\citep{Knight2012a}.  Such principles reduce the range
    of choices for design decisions and are also known as
    \emph{architectural tactics}~\citep{Wu2003}.
  \end{description}
  
\item[Procedures] %
  deal with the steps of the engineering process required to perform
  safety design and assurance.

  \begin{description}
  \item[Transformations] %
    cover the work steps of creating and changing engineering
    artifacts, \eg the hardening of an architecture, the corresponding
    update of a fault tree, or the refinement of a specification.
  
  \item[Arguments] %
    cover analysis and assessment practices to justify an acceptable
    level of safety of the considered system in a specific operational
    environment~\citep{Kelly1998,McDermid1994-Supportsafetycases}.
    Such arguments are used to establish \emph{claims}, \eg of the
    form ``System $S$ is free of hazard $H$,'' ``Development process
    of system $S$ complies with standard $X$,'' or ``Argument $A$ is
    sufficiently confident.''  The first one is also called a
    \emph{product-based} argument (\egs constructed during formal
    verification), the second one is a \emph{process-based} argument
    (in this case, a compliance argument), and the third one is a
    meta-argument (in this case, a confidence argument).
    \citet{Luo2016} provide a corresponding taxonomy.
  \end{description}
\end{description}
Such patterns reoccur in many of the collaborating branches of safety
engineering.  They are also part of widely used standards~(\egs
IEC~61508) and have received frequent attention in the scientific
literature.

\subsubsection{Terminology}

Subtle differences between a specification and a specification pattern
or between an argument and an argument pattern are not essential for
this survey.  However, reusable specifications and arguments should
be documented along with certain abstractions such as, \eg
type parameters, variation or extension points.
This survey is not about general software and hardware patterns.
For the sake of simplicity, we treat the terms hazard and (safety)
risk as synonyms.  Furthermore, we treat safety cases, or in general,
assurance cases as synonyms to assurance arguments.  However, an
argument is to be viewed distinctly from its representation, \eg a
goal structure that visualizes the argument.  According to
\citet{Hoare1985}, we view an implementation as a final and executable
refinement of a corresponding specification.

\subsubsection{From Designs to Specifications}
\label{sec:from-designs-spec}

Depending on the regulated domain, the engineering discipline, and the
level of abstraction, an enhancement leading to a
\emph{safety-enhanced design}~(\Cref{fig:overview}) is called ``risk
reduction measure'' [IEC~61508], ``safety function,'' ``safety
measure,'' ``safety pattern''~\citep{Preschern2013c}, ``safety
mechanism,'' ``safety-related system'' [IEC~61508], ``safety-related
function or element'' [ISO~26262], ``safety-critical
system''~\citep{Knight2012a}, or ``critical (computer)
system''~\citep{Rushby1994}.  \citeauthor{Rushby1994} and
\citeauthor{Knight2012a} adhere to the safety engineering tradition for
the description of these terms:
\begin{quotation}
  ``[A] critical (computer) system is a system whose malfunction could
  lead to unacceptable consequences.'' / ``[A] safety-critical system
  is a system whose consequences of failure are extremely serious.''
\end{quotation}
These terms suggest that only those systems that fail could engage in
undesired events.  However, \eg a mobile robot's behavior may well
cause loss of life without failure of its components, for instance, if
the robot lacks functionality to stop early for moving obstacles.%
\footnote{It now depends whether or not one counts a missing feature
  as a systematic fault.  Clearly, the robot in this example would of
  course no more comply with the state of the art.}
The definition of ``safety-critical service''
used by \citet{Burns1992-MeaningSafetySecurity} accommodates this
idea:
\begin{quotation}
  ``[A] service is judged to be safety-critical in a given context if
  its behavior could be sufficient to cause absolute harm to
  resources for which the enterprise operating the service has
  responsibility.''
\end{quotation}

\citet{Leveson2012} elaborates the idea of backwards reasoning from
accidents over hazards towards unsafe control actions.  Design choices
would then depend on whether such actions represent random or
systematic faults.  Her approach allows to focus on the \emph{system
  as operated in its context} able to engage in dangerous events (\ies
accidents) whether or not these events are caused by component
failure.
This way, \citeauthor{Leveson2012} supports the view of \emph{safety
  as a property} emerging from properties of elements of both the
system and its environment.  Given a system decomposition, safety can
be rephrased into a composite of component properties.
\citet{Rushby1994} discusses critical properties from several
viewpoints.  We will revisit these kinds of properties below.

Common to such properties is that their meanings depend on the
\emph{perimeter}, scope, boundary, or interface and the chosen
\emph{abstraction} they are specified for.  For hazard analysis, this
perimeter is usually congruent with the overall system. %
Consequently, a system safety property specifies behavior at the
system level.  

\subsubsection{Relationships between Patterns}
\label{sec:background-rel-pat}

The thick blue arcs in \Cref{fig:overview} depict three desirable
relationships among patterns of the mentioned categories:
\begin{description}
\item[Compositions and Behaviors express Designs] \Cref{fig:overview}
  already suggests that we take the view of a design having a
  behavioral and a structural
  facet~(\egs\citep{DBLP:series/natosec/Broy11}).  Both facets can be
  expressed in models.  If both models are given, we expect them to
  be consistent.
  
\item[Arguments and Transformations express Procedures] Arguments
  capture \emph{reasoning} steps from assurance evidence (\egs proofs,
  validation reports, test verdicts) towards assurance claims (\egs
  safety specifications).  Transformations capture \emph{engineering}
  steps.  We view arguments and transformations as procedures inasmuch
  as they incorporate work steps (\ies reasoning and construction
  steps) to be accomplished by engineers.

\item[Behaviors refine Specifications] If a system design is given in
  terms of a behavior, we want it to \emph{fulfill} the corresponding
  safety specification, in other words, we want it to be a
  \emph{refinement} of this specification.
  
\item[Compositions refine Behaviors refine Specifications] If a system
  design is given in terms of an architecture, a particular
  decomposition into components, two relationships are desirable:
  First, if a behavior is explicitly given, the behavior emerging from
  the composition should be consistent with (possibly, a refinement
  of) the explicitly given behavior.  Second, this composition should
  be refinement of the corresponding safety
  specification. \citet{DBLP:series/natosec/Broy11} gives a more
  comprehensive formal account of these relationships.

  After transforming a composition into another composition enhanced
  by principles such as, \eg recovery~(see, \egs\citep{Randell1975}),
  we desire that the mentioned relationships are maintained or
  established.
\end{description}

\subsubsection{Between Specifications and Arguments} %
\label{sec:background-behavior}

Given a specification and a design according to the left part of
\Cref{fig:overview}, safety engineers are interested in arguments for
two claims and measures if no arguments can be found for the current
versions of the artifacts (right part of \Cref{fig:overview}):
\begin{quotation}
  The specification does not imply relevant hazards.
\end{quotation}
If no argument can be found, the specification has to be transformed
into a safety specification ruling out all relevant hazards.  A safety
specification typically includes invariants requiring the system to
stay within safe regions~(see, \egs\citep{Leveson2012,Rushby1994}).
Such transformations can require the construction of a design as
discussed next.
\begin{quotation}
  The design is safe.  
\end{quotation}
This claim can be rephrased into the claim that the design is a
refinement of the safety specification.  Such a refinement holds if
and only if the weakest precondition for the design to fulfill the
safety specification is different from false.  However, if no such
argument can be found, the design has to be transformed into a
safety-enhanced design.  This transformation has to be consistently
performed if the design is given in terms of both artifacts.

\subsubsection{Between Specifications and Behaviors}
\label{sec:betw-spec-behav}

The behavioral perspective gives rise to \emph{behavioral tactics}
such as \emph{prevention} or \emph{active safety}, \emph{passive
  safety}, and \emph{fail-safe}~(\egs\citep{Knight2012a}).  The latter
represents the transition of a system to a safe state in the event of
failure.  The fail-safe tactic can be divided into the
\emph{fail-silent} tactic~(\ies transition to a safe state by
excluding failing components from the system functionality) and the
\emph{fail-operational} tactic~(\ies transition to a safe state
maintaining the original system functionality).

For any system and for each of its components, we can specify
dependability and security
properties~\citep{Laprie1992,Avizienis2004}.  As mentioned before,
safety can be seen as a property emerging from a system or to be
entailed by the compound behavior of its components.  This phenomenon
can be investigated using \emph{formal notions of properties} such as,
\eg safety, liveness, reliability, and availability.

\citet{Lamport1977-ProvingCorrectnessMultiprocess} formally discusses
two distinct behavioral properties of systems:
\begin{itemize}
\item \emph{Safety properties} state that something bad will never
  happen. %
\item \emph{Liveness properties} state that something good will eventually
  happen. %
\end{itemize}
\citet{Alpern1987-RecognizingSafetyLiveness} show that all behavioral
properties can be decomposed into a safety and a liveness part with
the obligation to proof invariance for safety and well-foundedness for
liveness of a specific system.
In concurrent systems with constrained resources, one also needs to
prove \emph{fairness} properties, \ie the property that each of a set
of components will infinitely often be able to be productive if they
wish to do so.  Hence, fairness is a special form of liveness.

\citet{Avizienis2004} qualitatively characterize \emph{reliability} as
the ``continuity of correct service.''  \citet{Knight2012a} and
\citet{Bertsche2009-ZuverlassigkeitmechatronischerSysteme} use two
related definitions of reliability: the ``probability that the system
will operate correctly [...] up until time $t$'' and the ``mean time
to the first (between two) failure(s).''
One can rephrase the latter
into the following
requirement:
\begin{quotation}
  The mean number of steps of system $S$ to the first ``bad thing''
  (between two ``bad things'') is greater than $n$.
\end{quotation}
If we substitute ``mean'' by ``minimum'' and let $n=\infty$, we get a
safety property.  The strength or weakness of a property corresponds
to the strength or weakness of its proof obligations.  Hence, the
requirement above suggests that proof obligations for reliability
requirements are in general weaker than proof obligations for formal
safety properties.  Consequently, reliability properties are in
general weaker than safety properties.

Although we can find qualitative abstractions of probabilistic
phenomena, reliability practitioners are usually interested in the
quantitative assessment of a system, particularly, in uncertainty
factors of technologies~(\egs material degradation and electromagnetic
interference causing random failures) and development processes~(\egs
developer mistakes causing systematic failures).
\Eg\citet{Littlewood2012} discuss stochastic process models to
calculate the \emph{probability of failure on demand} of a system with
a specific form of diverse redundancy.  Further reliability and
availability metrics include, \eg \emph{mean time between failure}.
Software defects can be considered as systematic if they are deployed
in the system as operated.  Hence, \citet{Littlewood1991} use
stochastic process models to predict the probability of development
defects occurring \emph{during} system use.

The introduced reliability definitions indicate the difference between
\emph{non-repairable} and \emph{repairable} systems.  Repairable
systems give rise to the discussion of \emph{availability}, \ie ``the
probability that the system will be operational at time $t$''
\citep[2.7.2]{Knight2012a} or its complement, the probability of
failure on demand.  From
\citeauthor{Lamport1977-ProvingCorrectnessMultiprocess}'s perspective,
a repairable system with non-zero availability would fulfill the two
liveness properties ``not always bad'' %
and ``always eventually good.''

Let us now look at the relationship of safety and
\emph{security}:
\citet{Avizienis2004} define \emph{security} as a composite of
\emph{integrity} (\ies the absence of unauthorized influence),
\emph{confidentiality} (\ies the absence of unauthorized access), and
\emph{availability} on demand of authorized actions.  Unauthorized
actions are the ``bad things'' that have to be reduced, or 
avoided if seen as a formal safety property.

\citet{Burns1992-MeaningSafetySecurity} distinguish between an
absolute and a relative ``degree of harm'' (also ``severity of
consequence'') to discriminate between safety
(absolute harm) and security (relative harm):
\begin{quotation}
  ``[A] service is judged to be security-critical in a given context if
  its behavior could be sufficient to cause relative harm, but never
  sufficient to cause absolute harm, to resources for which the
  enterprise operating the service has responsibility.''
\end{quotation}
The notions of
\citet{Burns1992-MeaningSafetySecurity} and \citet{Avizienis2004} can
be unified by a common principle:
\begin{quotation}
  The \emph{protection} of an \emph{asset} from an \emph{undesired
    event} caused by an \emph{agent}.
\end{quotation}
Proving the avoidance of the undesired event for a system
amounts to proving a formal safety property of this system.
For security, the assets would be represented by ``information'' or
``method calls'', the undesired event by ``unauthorized
access or influence through exploiting vulnerabilities,'' and the agent
by an ``attacker.''
For safety, the assets would be represented by ``humans'',
``animals'', or ``the environment,'' the undesired event by ``getting
harmed'', and the agent by ``the system under consideration.''
Below, we assume that designs can be specified with the discussed
properties.

\subsection{Objective of this Survey}
\label{sec:challenges}

We are interested in the characteristics of safety patterns and their
variety across several disciplines involved in safety engineering such as,
\eg software engineering, mechatronics, electrical engineering, 
mechanical engineering, and human factors engineering.  With this
survey, we aim at answering the question:
\begin{quote}
  What is the state-of-the-art of 
  design and %
  argument patterns %
  for the assurance of system safety?
\end{quote}
\Cref{tab:questions} decomposes this question into twelve survey
questions. 
By answering these questions, we aim at understanding various aspects
of patterns, the relationships between these patterns, their
properties and composition, and the relationships to their first
principles.

\begin{table*}[b]
  \caption{Overview of the survey questions}
  \label{tab:questions}
  \scriptsize
  \begin{tabularx}{\textwidth}{l>{\hsize=.4\hsize}X>{\hsize=.6\hsize}Xr}
    \toprule
    \textbf{Aspect} 
    & \textbf{Question}
    & \textbf{Expected Answer}
    & \textbf{Section}
    \\\midrule

    \RQ{4}{Application}
    & Which applications are discussed in the studies
      to demonstrate the patterns?
    & An overview of relevant studies by application domain
    & \ref{sec:rq-4-application}
    \\
    
    \RQ{4.1}{Mitigated Risk} %
    & Which types of risk are handled by the discussed patterns?
    & An overview of relevant studies by class and causal
      origin of risk
    & \ref{sec:rq-4.1-mitigation} 
    \\
    
    \RQ{3}{Engineering Step}
    & Which engineering steps are covered by the studies?
    & A classification into the categories: specification, %
      design, %
      implementation, %
      assurance %
    & \ref{sec:rq-3-engstep}
    \\

    \RQ{5}{Category}
    & Which categories of patterns are presented?
    & A classification into the categories: 
      specification,
      behavior,
      decomposition,
      procedure, 
      transformation,
      argument
    & \ref{sec:rq-5-cat}
    \\

    \RQ{1}{Pattern}
    & Which patterns are discussed?
    & An overview of the explained or applied patterns
    & \ref{sec:rq-1-patterns}
    \\

    \RQ{7}{Abstraction}
    & Which technologies are abstracted by the patterns?
    & A classification into the categories: software,
      electrical/electronic hardware, mechanical hardware 
    & \ref{sec:rq-7-abstraction}
    \\
    
    \RQ{6.1}{Tactic}
    & Which tactics are incorporated by the patterns?
    & A summary of the tactics covered according to the taxonomies in
      \citep{Preschern2013b} and
      \citep{Wu2003}
    & \ref{sec:rq-6.1-principle-dep}
    \\

    \RQ{12}{Relationship}
    & How are the patterns and tactics related to each other? %
    & An analysis based on the results of \RQRef{5}, \RQRef{1},
      \RQRef{6.1}
    & \ref{sec:rq12-patternrelationships}
    \\

    \RQ{6.2}{Contract}
    & Which behaviors are guaranteed by the patterns?
    & A summary of the system-level behaviors associated with the
      incorporated safety principles
    & \ref{sec:rq-6.2-principle-beh}
    \\

    \RQ{8}{Security}
    & How is security addressed in the studies?
    & A summary of how security is considered in the studies
    & \ref{sec:rq-8-secsafe}
    \\

    \RQ{2}{Model} %
    & Which models are used to describe the patterns?
    & An analysis and summary of the modeling 
      paradigms, formalisms, and notations
    & \ref{sec:rq-2-language}
    \\

    \RQ{15}{Contribution}
    & To what extent are known challenges covered by the studies?
    & A coverage analysis based on challenges from
      \citet{Cant2013,Langari2013,Graydon2017,Graydon2015} 
    & \ref{sec:rq15-challenges}
    \\
    \bottomrule	
  \end{tabularx}
\end{table*}

\subsection{Preliminary Work and Survey Method}
\label{sec:study_design}

For this survey, we first explored the field by an annotated
bibliography according to~\citet{Knott2015}.  Intermediate results are
reported in \citep{Gleirscher2016}.

For a more comprehensive overview of literature on safety patterns
and to determine research directions, we created a \emph{systematic map}
along the lines of \citet{DBLP:conf/ease/PetersenFMM08}.
We selected the most relevant studies by summed ranking of relevance.
For each selected study, we answered the survey questions in
\Cref{sec:challenges}.  Some questions~(\egs\RQRef{6.1}) involved 
qualitative content analysis~\citep{Neuendorf2016}, \ie content
abstraction by assignment of keywords to the studies and further
analysis based on these keywords.

\subsection{Related Work}
\label{sec:relwork}

In this section, we summarize literature studies on design and
argument patterns and studies of corresponding pattern taxonomies.

\paragraph{Literature Studies and Pattern Catalogs}

\citet{Preschern2014} discuss twelve safety-related pattern-based
methods regarding their target domain, the involved types of
patterns~(\ies process, safety tactics, architecture, timing), and the
degree of detail.  Their focus lies on pattern application and use in
the safety process. The authors state that they could not find a
similar study on safety patterns and their application.  We complement
their study with an extended analysis of pattern types and
relationships.

\citet{Luo2016} discuss a taxonomy of safety cases to map literature
on argument patterns and to establish a similarity relation
among safety cases.  In their work, four argument types are
distinguished: product-based arguments, process-based arguments,
compliance arguments, and confidence arguments.  The authors observe a
lack of research on confidence arguments.  Our work extends and
embeds their survey into a larger context.  Moreover, we use their
taxonomy below~(\cf\Cref{tab:rq5-cat-ap} in \Cref{sec:rq-1-patterns}).

\citet{Szczygielska2017} present an on-line argument pattern catalog
extracted from literature on argument patterns.  45 patterns were
modeled and can be instantiated from this catalog during the
construction of specific assurance cases.  While their catalog exceeds
the list of patterns we discuss in \Cref{tab:rq5-cat-ap}, our
contribution lies in establishing relationships among arguments and
between arguments and designs.

\citet{Langari2013} summarize challenges to be addressed by research
on assurance cases and to be taken into account in future standards
and regulations recommending assurance cases.  These challenges
include the identification of fallacious reasoning in arguments (\egs
confirmation bias), argument completeness, the specification of
assumptions, the reduction of argument size and complexity, and the
achievement of readability.  We extend \citeauthor{Langari2013}'s work
by identifying further research to solving these challenges.  We
explore the state of the art, how far research has come in addressing
such challenges, as well as interesting research gaps.

\citet{Havarneanu2015} summarize behavioral safety tactics resulting
from accident research in the railway domain, viewing railways as a
socio-technical system~\citep{Leveson2012}.  The authors summarize
measures for the prevention of trespass accidents, \eg barriers,
organizational measures, monitoring, enforcement, track design, staff
training, station lighting, and rail traffic management.  Our study
takes an engineering perspective on designs and arguments and,
moreover, abstracts from prevention measures of a specific domain.

\citet{Kakamanshadi2015} summarize research on fault-tolerance
mechanisms for resilient and reliable wireless sensor networks.  These
mechanisms are based on redundancy of nodes, paths, data, and time; on
clustering to reduce performance bottlenecks, and on optimal
deployment.  Taking a more general view, our survey can help to bridge
the gap between network reliability and performance and the concepts
required in safety-critical applications relying on such networks.

\paragraph{Tactic Taxonomies}

\citet{Kumar2010} provide a framework for developing taxonomies such
as the ones for safety patterns discussed by \citet{Wu2003} and
\citet{Preschern2013c}.
\citet{Ryoo2012} describe the extraction and revision of tactic
hierarchies for the consolidation of pattern catalogs.  Their
application to security tactics provides insight on how safety and
security patterns could be aligned on a tactics level.
\citet{Hawkins2012} show guidance on the construction of safety
arguments, enumerating principles helpful for the choice of argument
patterns.
Our analyses in the \Cref{sec:rq-1-patterns,sec:rq-6.1-principle-dep} are
based on these works.

\subsection{Contributions and Outline}
\label{sec:contrib}

We present a survey of reusable concepts for the specification,
design, and assurance of safety-critical systems.  Based on the survey
questions presented in \Cref{tab:questions}, we classify a range of 
studies of such concepts by their application domains, type of
mitigated risk, the supported engineering steps, their first
principles, and their abstractions.  From a cross-disciplinary
perspective, this survey provides information 
\begin{itemize}
\item to identify relevant contributions to this field,
\item about the range of safety-enhanced designs and
  assurance arguments,
\item to develop a unified view of reusable designs and arguments, and
\item to identify directions for further research.
\end{itemize}
With this study, we contribute to the consolidation of the practical
safety engineering body of knowledge.
Based on preliminary materials in~\citep{Gleirscher2016} and on the
systematic map according to
\citep{DBLP:conf/ease/PetersenFMM08}~(\Cref{sec:study_design}%
), this survey forms a
\emph{systematic literature review} according
to~\citet{Kitchenham2007}.

The rest of this work is structured as follows: \Cref{sec:results}
presents the answers to the survey questions.  \Cref{sec:discussion}
discusses these answers, derives recommendations for future
research~(\Cref{sec:findings}), and discloses limitations of our
study~(\Cref{sec:threats}). We draw final conclusions in
\Cref{sec:conclusion}.

\section{Aspects of Design and Argument Patterns for Safety Assurance}
\label{sec:results}

The following sections provide answers to the survey questions as
introduced in \Cref{tab:questions}.

\subsection{\RQRef{4}: Which applications are discussed to demonstrate
  the patterns?}
\label{sec:rq-4-application}

\Cref{tab:rq4} lists the most relevant studies by application domain.
The surveyed studies cover application domains such as, \eg process
plants, machinery, automotive, and avionic systems.  Many studies
present generically applicable patterns.
Across the mentioned domains, our survey focuses on pattern
applications for the design and assurance of embedded control systems
and distributed systems.

\begin{table}
  \caption{\RQRef{4}: Selection of studies by application domain
    \label{tab:rq4}}
  \footnotesize
  \begin{tabularx}{1.0\linewidth}{L{.2}L{.8}}
    \toprule
    \textbf{Application Domain}
    & \textbf{Selection of Studies}
    \\\midrule
    Aircraft \& avionics
    &  \cite{Gobbo2001, Basir2010, Delange2009, Delmas2015,
      Denney2013, Dias2011, Kehren2004, Kelly2006, Lopez-Jaquero2012,
      Miller2009, Mueller2012, Netkachova2015, Steiner2011a,
      Littlewood2012, Zeng2016, Delmas2017}  
    \\
    Automotive
    & \cite{Dardar2012, Domis2009, Palin2010, Trindade2014, Wu2003,
      Luo2017, Nasser2017, Antonino2014, Ebnenasir2007, Gallina2014,
      Hocking2014, Konrad2004, Oertel2014a, Oliveira2015a, Owda2015,
      Penha2015a, Pont2003, Rupanov2012, Sljivo2015, Standish2014a,
      Wagner2010, Lin2016, Martorell2016}  
    \\
    Machinery \& railway
    & \cite{Eloranta2010, Orlic2007, Tan2009, Sljivo2017,
      Preschern2013a, Gleirscher2017, Hauge2013, Radermacher2013, Havarneanu2015} 
    \\
    Medical \& other devices; healthcare
    & \cite{Lin2015, Sun2010a, Sun2014, Tan2015, Fayad2003, Pont2004,
      Lakhani2012, Murugesan2015, Preschern2014a} 
    \\
    Networks \& telecommunication
    & \cite{Saridakis2002, Saridakis2003, Giuntini2017, Petroulakis2016} 
    \\
    Process \& power plants    
    & \cite{Mahemoff2001, Rauhamaeki2012, Rauhamaeki2015, Larrucea2016,
      Wilson1992, Larrucea2017}
    \\\midrule
    Generic
    & \cite{Alho2011, Armoush2008, Armoush2008a, Armoush2009a,
      Armoush2010, Baleani2003, Bozzano2013, Crenshaw2006,
      Denney2013a, Grunske2003, Habli2010a, Islam1996, Kabir2015,
      Kelly1998, Mayo2006, Natarajan2000, Preschern2013b,
      Preschern2013c, Preschern2014, Weaver2003, Knight2012a,
      Randell1975, Luo2016, Chen2016, Iliasov2008, Liu2008} 
    \\\bottomrule
  \end{tabularx}
\end{table}

\subsection{\RQRef{4.1}: Which types of risk are handled by the
  discussed patterns?}
\label{sec:rq-4.1-mitigation}

\begin{table}
  \caption{\RQRef{4.1}: Selection of studies by type and causal origin of risk
    \label{tab:rq4.1}}
  \footnotesize
  \begin{tabularx}{1.0\linewidth}{L{.2}L{.8}}
    \toprule
    \textbf{Risk Class}
    & \textbf{Selection of Studies}
    \\\midrule
    Specification \& design defects (systematic) %
    & \cite{Konrad2004, Sljivo2015, Miller2009, Mahemoff2001}
    \\ 
    Behavioral hazards \& accidents
    & \cite{Lin2015, Tan2009, Murugesan2015, Napolano2015,
      Rauhamaeki2014, Riera2014, Sun2010a, Havarneanu2015} 
    \\ %
    Generic technical defects (mainly random, in hardware)
    & \cite{Armoush2008, Armoush2010, Iliasov2008, Kehren2004,
      Orlic2007, Preschern2013b, Preschern2013c, Preschern2014,
      Rauhamaeki2012, Rauhamaeki2015, Saridakis2002, Littlewood2012,
      Randell1975, Chen2016, Delmas2017, Mueller2012, Baleani2003,
      Grunske2003, Tan2015, Knight2012a} 
    \\
    High complexity 
    & \cite{Kelly1998, Kelly2006, Miller2009, Palin2010, Sljivo2017,
      Luo2016, Ebnenasir2007, Jackson2001, Lakhani2012,
      Ljungkrantz2012, Radermacher2013, Pont2001, Lin2016,
      Larrucea2017, Martorell2016, Sorokos2016} 
    \\
    Undesired interference \& mixed criticality %
    & \cite{Larrucea2015a, Owda2015, Rauhamaeki2011, Netkachova2015,
      Kehren2004, Orlic2007, Larrucea2017, Larrucea2016, Tan2015, Althammer2008a} 
    \\ 
    Security threats
    & \cite{Castellanos2013, Preschern2013a, Petroulakis2016,
      Nasser2017, Cimatti2015, Netkachova2015}
    \\
    Argument flaws
    & \cite{Gleirscher2017, Mayo2006, Rich2007, Standish2014a}
    \\ %
    \midrule
    At system-level (from several technologies)
    & \cite{Gleirscher2017, Lin2015, Tan2009, Luo2016, Crenshaw2006,
      Preschern2013a, Nasser2017, Armoush2010, Gobbo2001, Kehren2004,
      Lopez-Jaquero2012, Preschern2013b, Preschern2013c,
      Preschern2014, Rauhamaeki2012, Rauhamaeki2015, Littlewood2012,
      Randell1975, Zeng2016, Delmas2017, Mueller2012, Baleani2003,
      Fayad2003, Grunske2003, Sun2014, Knight2012a, Sun2010a}
    \\ %
    Purely from software
    & \cite{Natarajan2000, Pont2004, Armoush2008, Armoush2008a,
      Iliasov2008, Saridakis2002, Chen2016}
    \\
    Purely from hardware
    & \cite{Steiner2011a, Petroulakis2016, Delmas2015, Islam1996,
      Kabir2015, Giuntini2017, Luo2017} 
    \\
    \bottomrule
  \end{tabularx}
\end{table}

\Cref{tab:rq4.1} is organized according to frequently discussed types
of risk including its causal origin or location of occurrence.  The
surveyed studies deal with the reduction of various technical defects
and the reduction of complexity.  The presented approaches handle such
risks by measures built from several technologies (\ies software,
electronic hardware, mechanical hardware).  The investigations deal
with risks stemming, \eg from the system as a whole, purely from
software, purely from hardware, from requirements and system design,
from arguments, from undesired interference.

\subsection{\RQRef{3}: Which engineering steps are covered by the
  studies?}
\label{sec:rq-3-engstep}

In \Cref{sec:engineeringsteps}, we distinguish the four engineering
steps of specification, design, implementation, and assurance.  Most
of our work focuses on studies allocated to the design and assurance
steps of this process.  Several studies cover at least three steps and
describe the transitions between them.

Two studies bridge the gap between \emph{specification and
  implementation}:
\citet{Oertel2014a} note that conventional arguments that an
implementation fulfills a safety specification are often based on
matching fault trees and test results with this specification.  The
authors employ fault-injection and model checking against safety
contracts and, this way, cover the four considered engineering steps.
\citet{Trindade2014} show how formalized safety requirements
help transforming a system design into a new design enhanced with
safety mechanisms.

Several studies discuss the transition from \emph{specification over
  design to assurance}:
\citet{Gobbo2001} describe a refinement-based approach to deriving
fault-tolerant specifications of flight control systems.  Assurance is
achieved by maintaining relational refinement across specification and
design steps.
\citet[Ch.~6 and 11]{Knight2012a} provides a comprehensive treatment
of  the specification, design, and
assurance steps of the
software dependability life cycle.  Particularly, he discusses 
designs (\egs N-modular redundancy) reducing the
negative impact of degradation faults on software dependability, %
designs (\egs recovery blocks) and procedures (\egs
N-version programming) for improving software fault-tolerance, %
and
the creation of rigorous arguments. %
Further discussions that relate specification, design, and assurance
are provided by \citet{Antonino2014}, \citet{Cimatti2015},
\citet{Domis2009}, \citet{Hauge2013}, \citet{Sorokos2016},
\citet{Wilson1992}, and \citet{Wu2013}.

\subsection{\RQRef{5}: Which categories of patterns are presented?}
\label{sec:rq-5-cat}

Among the categories introduced in \Cref{sec:safetypatterns}, our work
concentrates on designs and arguments.  However, more than half of the
studies cover at least two of the considered categories.
\Cref{tab:rq5-selections} lists studies representing these categories.

\begin{table}
  \caption{\RQRef{5}: Selections of studies by pattern category
    \label{tab:rq5-selections}}
  \footnotesize
  \begin{tabularx}{1.0\linewidth}{lX}
    \toprule
    \textbf{Category}
    & \textbf{Selection of Studies}
    \\\midrule
    Specification
    & \citet{Basir2010,Gobbo2001,Kehren2004,Trindade2014,Jackson2001,
      Kajtazovic2014, Konrad2004, Oertel2014a, Sljivo2015} 
    \\
    Behavior
    & \citet{Baleani2003,Basir2010,Crenshaw2006,Mahemoff2001,%
      Mueller2012,Palin2010,Rauhamaeki2012,Rauhamaeki2015,%
      Sun2010a,Tan2009,Belli1991, Havarneanu2015}
    \\
    Decomposition
    & \citet{Armoush2008, Armoush2010, Baleani2003, Crenshaw2006,
      Delange2009, Dias2011, Domis2009, Grunske2003, Iliasov2008,
      Kehren2004, Liu2008, Miller2009, Mueller2012, Orlic2007,
      Pont2004, Preschern2013a, Preschern2013b, Preschern2013c,
      Preschern2014, Rauhamaeki2012, Rauhamaeki2015, Saridakis2002,
      Sun2010a, Tan2015, Trindade2014, Wu2003, Knight2012a,
      Littlewood2012, Randell1975, Chen2016, Delmas2017} 
    \\
    Transformation
    & \citet{Delmas2015, Domis2009, Grunske2003, Kehren2004,
      Trindade2014, Sljivo2017, Delmas2017, Castellanos2013,
      Getir2018-Supportingsemiautomatic}  
    \\
    Argument
    & \citet{Basir2010, Bozzano2013, Dardar2012, Denney2013,
      Denney2013a, Domis2009, Habli2010a, Kelly1998, Kelly2006,
      Lin2015, Mayo2006, Netkachova2015, Palin2010, Preschern2013c,
      Sun2010a, Weaver2003, Wu2003, Knight2012a, Sljivo2017, Luo2016,
      Larrucea2016, Gleirscher2017, Chen2016} 
    \\
    Procedure
    & \citet{Bozzano2013, Denney2013, Domis2009, Kehren2004,
      Kelly2006, Lin2015, Mayo2006, Netkachova2015, Palin2010,
      Preschern2014} 
    \\\bottomrule
  \end{tabularx}
\end{table}

We found two works that cover \emph{four of the categories}.
\citet{Kehren2004} present a state-machine based approach to the
generic modeling of safety functions for the refinement of a class of
component architectures.  They discuss the automatic proof of
properties of these functions and, by refinement, of all architectures
enhanced by these functions, using a linear temporal logic model
checker.  The authors demonstrate their method with an architecture
using cold redundancy. %
\citet{Domis2009} show a concept for achieving traceability of safety
information (\ies argumentation evidence) throughout a component-based
safety engineering life cycle.  Component fault trees derived from
hazard analysis provide the core structure for deriving safety
requirements (\egs by hazard negation) and for constructing the safety
argument by modifying the original fault trees.  This approach is
exemplified by integrating a safety limiter into an automotive braking
controller.

Notable are also the following studies covering \emph{three of the
  categories}:
Using a railway interlocking system as an example, \citet{Hauge2013}
explain a method for the systematic development of safety concepts.
For demonstration, the authors describe the integration of redundancy
into this system.  During typical safety engineering steps (\ies
elicit functional requirements, elicit safety requirements, establish
design basis, establish safety case), appropriate safety patterns are
selected, instantiated, and composed.  The patterns instantiated
throughout these steps are then synthesized into a ``composite pattern
solution.''
Given a library of safety mechanisms and a software safety
requirement, \citet{Trindade2014} show how a given software
implementation can be automatically enhanced by a safety mechanism
such that the safety requirement is fulfilled.
To increase the level of reuse in safety-critical systems engineering,
\citet{Kajtazovic2014} propose a structure for a safety pattern
database using contract methodology~\citep{Meyer1992}.

Several studies bridge the gap between \emph{two categories}:
\citet{Gobbo2001} show how requirements specifications can
be formally refined into conceptual designs.
For safety-related design tactics, \citet{Grunske2003} presents a
catalog of architectural transformations with the goal of hardening
software architectures according to these tactics.
Given an architectural decomposition, \citet{Delmas2015,Delmas2017}
automate the identification of enhancements of this decomposition to
meet safety requirements.  Along the lines of \citet{Kehren2004},
these enhancements incorporate design principles (\egs redundancy)
used for ``hardening'' of an architecture. %
The authors identify parameters forming a design space and use SMT
solvers to find best solutions to the corresponding constraint
satisfaction problem.
Similar to the approach of \citet{Domis2009},
\citet{Getir2018-Supportingsemiautomatic} demonstrate how
corresponding \emph{architecture/fault-tree pairs} can be
automatically co-evolved---for a specific set of modeling
operations---using coupled model transformations.
\citeauthor{Getir2018-Supportingsemiautomatic}'s approach can be
combined with transformations for architecture hardening as discussed
by \citet{Grunske2003}.

The relationship between specifications and
arguments~(\Cref{sec:background-rel-pat}) is discussed in
\cite{Antonino2014, Basir2010, Jaradat2015, Kajtazovic2014,
  Kotonya1997, Sljivo2015}.
Both aspects of designs, behaviors and decompositions, receive a
treatment in the works of \citet{Mahemoff2001, Mueller2012,
  Rauhamaeki2012, Rauhamaeki2015, Sun2010a}.
The transition between a system decomposition and the construction of
a safety argument is taken account of by \citet{Netkachova2015,
  Preschern2013c, Sun2010a, Wu2003, Knight2012a, Larrucea2016,
  Chen2016}.
Several authors discuss both arguments and procedures in a reusable context
\cite{Kelly2006, Lin2015, Mayo2006, Netkachova2015, Palin2010}.

\subsection{\RQRef{1}: Which patterns are discussed?}
\label{sec:rq-1-patterns}

The \Cref{tab:rq5-cat-ap,tab:rq5-cat-dp} relate frequently discussed
specifications, arguments, and designs %
with the tactics %
they rely on.  The tables contain information about the
abstraction %
and whether the pattern is associated with a behavioral
constraint %
as explained in \Cref{sec:background}.
We also indicate relationships %
such as \emph{generalizes} in \Cref{tab:rq5-cat-dp} and
\emph{supported-by} in \Cref{tab:rq5-cat-ap}.  We do not focus cases
where one pattern uses another one, \eg if a distributed architecture
uses a safety kernel.  However, readers interested in further details
about these and further relationships may consult the references given
in the tables, particularly,
\citet{Palin2010,Preschern2013c,Rauhamaeki2012}.

\begin{table*}
  \caption{%
    Specification and design patterns (behaviors and decompositions)}
  \label{tab:rq5-cat-dp}
  \scriptsize
  \begin{tabularx}{\textwidth}{L{.4}|lccc|ccccccccc}
    \textbf{Specification/Design (\RQRef{1})}
    & \textbf{Refs.}
    & \rb{\RQRef{7}} %
    & \rb{Type \cite{Luo2016}}
    & \rb{Behavioral constraints} %
    & \rb{Preventive/passive} %
    & \rb{Fail-safe, fail-over} %
    & \rb{Fault avoidance} %
    &   \rb{Checking} %
    &   \rb{Comparison}
    &   \rb{Redundancy} %
    &   \rb{Recovery} %
    &   \rb{Masking, limiter} %
    & \rb{Barrier, separation} %
    \\\midrule
    
    Safety contract  
    & \MetaciteRef{Oertel2014a} \MetaciteRef{Basir2010} \MetaciteRef{Sljivo2017}
    &S&pd&*& & & &M \\ %
    Safety concept (de)composition  
    & \MetaciteRef{Antonino2014} \MetaciteRef{Antonino2015}
      \MetaciteRef{Basir2010} \MetaciteRef{Domis2009}
    &S& &*& & &*& & &* \\ %
    $\hookrightarrow$ Parametric safety concept spec. 
    & \MetaciteRef{Antonino2014}
    &S& & & & & & & &* \\
    \midrule

    Hardware platform reassignment
    & \MetaciteRef{Grunske2003}
    &H& & & & &Su\\
    Hardware platform substitution
    & \MetaciteRef{Grunske2003}
    &H& & & & &Su\\
    Process fusion
    & \MetaciteRef{Grunske2003}
    &S& & & & &Si\\
    Distributed multiple indep. levels of security
    & \MetaciteRef{Cimatti2015} \MetaciteRef{Althammer2008a}
    &*&pd&C& &fs&Si&M& & & &O&S\\
    Safety kernel
    & \MetaciteRef{Wu2011}  %
    &*&pd& & & &Si& & & & & &S\\
    Separated safety                     
    & \MetaciteRef{Rauhamaeki2012}
    &*& & & &*&Si& & & & & &S\\
    $ \hookrightarrow$ Productive safety  
    & \MetaciteRef{Rauhamaeki2012}
    &*& & & & & & & \\
    $\quad \hookrightarrow$ Hardwired safety      
    & \MetaciteRef{Rauhamaeki2012}
    &H& & & & & & & \\
    $ \hookrightarrow$ Separated override
    & \MetaciteRef{Rauhamaeki2012}
    &H& & & & & & & & & &O\\
    $ \hookrightarrow$ De-energized override          
    & \MetaciteRef{Rauhamaeki2012}
    &H& & & & & & & & & &O\\
    $ \hookrightarrow$ Safety limiter             
    & \MetaciteRef{Rauhamaeki2012}
    &H& & &*& & & & & & &*\\

    M-out-of-N, multi-channel redundancy %
    & \MetaciteRef{Knight2012a} \MetaciteRef{Grunske2003} \MetaciteRef{Preschern2013c} \MetaciteRef{Preschern2013a}
    &*& & & &fo& & & &R& &V\\    	
    
    $ \hookrightarrow$ Triple modular redundancy     
    & \MetaciteRef{Grunske2003} \MetaciteRef{Preschern2013c} \MetaciteRef{Preschern2013a}
    &H& & & &fo\\
    
    $ \hookrightarrow$ M-out-of-N-D         
    & \MetaciteRef{Preschern2013c} \MetaciteRef{Preschern2013a}
      \MetaciteRef{Littlewood2012}
    &*& & & &fo& &M& & & &O\\ 
    $\quad \hookrightarrow$ Homogeneous duplex, 2-ch. red.
    & \MetaciteRef{Grunske2003} \MetaciteRef{Preschern2013c} \MetaciteRef{Preschern2013a}
    &H& & & &fo \\ %
    $\quad \hookrightarrow$ Heterogeneous duplex      
    & \MetaciteRef{Grunske2003} \MetaciteRef{Preschern2013c}
      \MetaciteRef{Preschern2013a}
    &H& & & &fo& & & &D\\
    $\quad \hookrightarrow$ Recovery block      
    & \MetaciteRef{Grunske2003} \MetaciteRef{Preschern2013c} \MetaciteRef{Preschern2013a}
    &S& & & &fo& &S& &*&R\\
    $ \hookrightarrow$ N-version-programming          
    & \MetaciteRef{Preschern2013c} \MetaciteRef{Preschern2013a}
    &S& & & &fo& & & &D\\
    $\quad \hookrightarrow$ Acceptance voting          
    & \MetaciteRef{Preschern2013c} \MetaciteRef{Preschern2013a}
    &S& & & &fo& &S& &D& &V \\
    $\quad \hookrightarrow$ N-self checking programming          
    & \MetaciteRef{Preschern2013c} \MetaciteRef{Preschern2013a}
    &S& & & &fo& & &*&D&D\\

    Actuation monitor %
    & \MetaciteRef{Grunske2003} \MetaciteRef{Preschern2013c}
      \MetaciteRef{Preschern2013a} \MetaciteRef{Rauhamaeki2012}
    &H& & & &fs& &*\\
    $ \hookrightarrow$ Watchdog, sanity/integrity check %
    & \MetaciteRef{Grunske2003} \MetaciteRef{Preschern2013c}
      \MetaciteRef{Preschern2013a}
    &H& & & &fs& &S& & & &O\\
    $\quad \hookrightarrow$ 3-level safety monitoring     
    & \MetaciteRef{Preschern2013c} \MetaciteRef{Preschern2013a}
    &*& & & &fs& &*& & & &O\\
    $ \hookrightarrow$ Protected single channel      
    & \MetaciteRef{Grunske2003} \MetaciteRef{Preschern2013c}
      \MetaciteRef{Preschern2013a}
    &*& & & &fs& &*& & & &O\\

    Safety executive
    & \MetaciteRef{Preschern2013c} \MetaciteRef{Preschern2013a}
    &H& & & &*& &S& & &D&O\\
    \bottomrule
    \multicolumn{14}{X}{
    \textbf{Legend:}
    $\hookrightarrow$\dots\emph{generalizes},
    * \dots generic,
    (S)oftware, (H)ardware,
    fo \dots fail-over/operational,
    fs \dots fail-safe/silent,
    pd \dots product-based argument,
    pr \dots process-based argument,
    cm \dots compliance argument,
    cf \dots confidence argument,
    (C)ontract,
    (Su)ubstitution,
    (Si)mplicity,
    (M)onitoring,
    (S)anity check,
    (R)eplication,
    (D)iversity,
    (R)epair,
    (D)egradation,
    (O)verride,
    (V)oting.\newline
    Discussion in, \egs
    \Metacite{Althammer2008a},
    \Metacite{Antonino2014},
    \Metacite{Antonino2015},
    \Metacite{Basir2010},
    \Metacite{Cimatti2015},
    \Metacite{Domis2009},
    \Metacite{Grunske2003},
    \Metacite{Knight2012a},
    \Metacite{Littlewood2012},
    \Metacite{Oertel2014a},
    \Metacite{Preschern2013a},
    \Metacite{Preschern2013c},
    \Metacite{Rauhamaeki2012},
    \Metacite{Sljivo2017},
    \Metacite{Wu2011}
    }
  \end{tabularx}
\end{table*}

\begin{table*}
  \caption{%
    Argument patterns
  \label{tab:rq5-cat-ap}}
  \scriptsize
  \begin{tabularx}{\textwidth}{L{.3}|lcccc|ccccccccc}
    \textbf{Argument (\RQRef{1})}
    & \textbf{References}
    & \rb{\RQRef{7}} %
    & \rb{Type \cite{Luo2016}}
    & \rb{Hierarchy/modularity} %
    & \rb{Behavioral constraints}%
    & \rb{Preventive/passive} %
    & \rb{Fail-safe, fail-over} %
    & \rb{Fault avoidance} %
    &   \rb{Checking} %
    &   \rb{Comparison}
    &   \rb{Redundancy} %
    &   \rb{Recovery} %
    &   \rb{Masking, limiter} %
    \\\midrule
    Generic modules
    & \MetaciteRef{Denney2015b} \MetaciteRef{Despotou2008}
      \MetaciteRef{Kelly2006} \MetaciteRef{Sljivo2017} \MetaciteRef{Althammer2008a}
    &*&*&\checkmark& &  \\
    Requirements~(de)composition %
    & \MetaciteRef{Basir2010} \MetaciteRef{Hawkins2009}
      \MetaciteRef{Palin2010} \MetaciteRef{Antonino2015} %
    &S&pd&\checkmark&*& & & &*& & \\
    $\hookrightarrow$ Requirements formalization %
    & \MetaciteRef{Basir2010}
    &S&pd& &*& & & &*& & \\
    $\quad \hookrightarrow$ Property-oriented %
    & \MetaciteRef{Basir2010}
    &S&pd& &* \\
    $\quad\quad \hookrightarrow$ Safety notion %
    & \MetaciteRef{Basir2010}
    &S&pd& &* \\    
    $\hookrightarrow$ Calculate/convey/use
    & \MetaciteRef{Feather2011}
    &S&pd& & & & & &*& & & &* \\
    $\hookrightarrow$ Interface safety (human-factors)
    & \MetaciteRef{Rich2007}
    &*&pd& & &* \\
    Configurable architectures (prod. lines)
    & \MetaciteRef{Habli2010a}
    &S&pd&\checkmark& &  \\
    Predefined safety requirements 
    & \MetaciteRef{Palin2010}
    &*&cm,cf& & &* \\
    $\hookrightarrow$ Homologation or backing
    & \MetaciteRef{Palin2010}
    &*&pr& & &  \\
    Process compliance
    & \MetaciteRef{Gallina2014}
    &S&cm,pr& & &  \\
    Product compliance
    & \MetaciteRef{Larrucea2016} \MetaciteRef{Larrucea2017}
    &*&cm,pd&\checkmark& &  \\
    High-level vehicle safety
    & \MetaciteRef{Palin2010}
    &*&*& & & & & & & &*&*&* \\ 
    $ \hookrightarrow$ High-level SW safety
    & \MetaciteRef{Hawkins2009}
    &S&pd& & & & & & & &*& \\ 
    $\quad \hookrightarrow$ SW contribution safety
    & \MetaciteRef{Hawkins2009} \MetaciteRef{Weaver2003}
    &S&pd& & & & &*& & &*&*&* \\ 
    $\quad \hookrightarrow$ SW safety requirements    
    & \MetaciteRef{Hawkins2009} \MetaciteRef{Antonino2014} \MetaciteRef{Antonino2015}
    &S&pd& & & & &*& & &*& &L \\ 
    $\quad\quad \hookrightarrow$ Argument justification SW
    & \MetaciteRef{Hawkins2009}
    &S&cf& & & \\ %
    Risk management                
    & \MetaciteRef{Palin2010} \MetaciteRef{Gleirscher2017}
    &*&pr&\checkmark& & \\
    $ \hookrightarrow$ Safety goal valid              
    & \MetaciteRef{Palin2010}
    &*&cf& & & & &*\\
    $ \hookrightarrow$ Minimization
    & \MetaciteRef{Palin2010}
    &*& & & &* \\
    $\quad \hookrightarrow$ Alert and warning              
    & \MetaciteRef{Palin2010}
    &*& & & &pr \\
    $ \hookrightarrow$ Hazardous contrib.\ SW, risk mitig. 
    & \MetaciteRef{Hawkins2009} \MetaciteRef{Palin2010} %
      \MetaciteRef{Antonino2014}
      \MetaciteRef{Antonino2015}
      \MetaciteRef{Sorokos2016} \MetaciteRef{Yuan2010} \MetaciteRef{Lin2015}
    &*& & & &*&*&*&*&*&*&D&* \\
    $ \hookrightarrow$ Hazard identification          
    & \MetaciteRef{Palin2010} \MetaciteRef{Gleirscher2017}
    &*&pr&\checkmark& & & \\
    $\quad \hookrightarrow$ Failure-mode-effects analysis
    & \MetaciteRef{Palin2010} \MetaciteRef{Alexander2008a}
    &*&pr& & & &  \\
    Risk assessment                
    & \MetaciteRef{Palin2010}
    &*&pr& & & &  \\
    $\hookrightarrow$ Safety assessment model adequate
    & \MetaciteRef{Sun2011}
    &*&cf& & & & \\
    Product defects, production errors               
    & \MetaciteRef{Palin2010} \MetaciteRef{Preschern2013c}
    &*& & & & & &* \\ %
    Through life safety            
    & \MetaciteRef{Palin2010}
    &*&pr& & &pr \\ %
    \bottomrule
    \multicolumn{15}{X}{
    \textbf{Legend:} $\hookrightarrow$\dots\emph{supported-by},
    similar patterns in the same row,
    see \Cref{tab:rq5-cat-dp},
    pr \dots preventive,
    D \dots degradation,
    L \dots limiter.
    \newline
    Discussion in, \egs
    see \Cref{tab:rq5-cat-dp},
    \Metacite{Alexander2008a},
    \Metacite{Denney2015b},
    \Metacite{Despotou2008},
    \Metacite{Feather2011},
    \Metacite{Gallina2014},
    \Metacite{Gleirscher2017},
    \Metacite{Habli2010a},
    \Metacite{Hawkins2009},
    \Metacite{Kelly2006},
    \Metacite{Larrucea2016},
    \Metacite{Larrucea2017},
    \Metacite{Lin2015},
    \Metacite{Palin2010},
    \Metacite{Rich2007},
      \Metacite{Sorokos2016},
      \Metacite{Sun2011},
    \Metacite{Weaver2003},
    \Metacite{Yuan2010}
    }
  \end{tabularx}
\end{table*}

The following analysis addresses the questions 
\RQRef{3},
\RQRef{7}, 
\RQRef{5}, and 
\RQRef{2} 
from 
\Cref{tab:questions}.
The column ``References'' in the \Cref{tab:rq5-cat-dp,tab:rq5-cat-ap}
recommends works providing more detailed explanations of the listed
patterns.

\paragraph{Specifications}

\citet{Antonino2014} discuss the issue of inconsistencies between
safety concepts and architecture designs.  It is difficult to keep
assurance artifacts~(\egs as defined in ISO~26262) up to date and
safety concepts consistent with an evolving architecture.  The authors
propose (i) a \emph{safety concept decomposition pattern} and (ii)
\emph{parametrized safety concept specification templates}.  An
example of a power sliding door module illustrates their approach.

\citet{Antonino2015} propose a procedural pattern for
decomposing safety requirements such that traceability to an
architectural design and a fault propagation model~(\ies fault
  trees) is established to perform complete and consistent
hazard %
mitigation.  The authors describe traceability between safety
requirements, functional and technical architecture, and fault trees.

\citet{Oertel2014a} investigate checking of safety requirements using
formalized contracts~(\Cref{sec:background}) expressed through
property patterns translated into LTL and applied in the VIS checker
for MatLab/Stateflow models.  The authors determine fault
combinations, injected into these models, resulting in a contract
violation and demonstrate this idea for an automotive light manager.

\paragraph{Decompositions (and Transformations)}

For a given decomposition, \citet{Grunske2003} proposes
transformations for many of the software and hardware patterns listed
in \Cref{tab:rq5-cat-dp}.  These transformations can be applied to a
decomposition resulting in a safety-enhanced decomposition
implementing one or more of the principles, \eg substitution,
checking, redundancy, recovery.  These principles aim at improving
safety, \ie by reducing hazard probabilities.  The transformations aim
at handling three types of component faults: unavailability, faulty
reactions, and timing deviations.  \citet{Grunske2003} discusses
patterns for fault avoidance and for fault
containment~(\Cref{tab:rq5-cat-dp}).  He proposes to refine these
patterns with patterns for fault detection~(\ies watchdog, integrity
check, and the actuation monitor, see also \citep{Rauhamaeki2012}).

\citet{Rauhamaeki2012} discuss hardware and software design patterns
for control and safety system development.
They describe \emph{separated safety} as their main pattern as well as
\emph{productive safety}~(a pattern for high-level safe system
behavior), \emph{separated override}, \emph{de-energized override},
\emph{safety limiter}~(a pattern for preventive safety actions), and
\emph{hardwired safety}.  Each of the latter refines the
\emph{separated safety} pattern~(see \Cref{tab:rq5-cat-dp}).
\citeauthor{Rauhamaeki2012} sketch structural and behavioral details
of their patterns.
Of particular interest is their interdisciplinary discussion to
capture reusable knowledge beyond the domain of software design.

\citet{Preschern2013c} summarize decomposition patterns of
fault-tolerant systems in safety-critical applications.  They identify
\emph{safety tactics}~(\Cref{sec:rq-6.1-principle-dep}) underlying
each pattern, construct product-based safety arguments to understand
how the pattern implements these tactics (\egs replication
redundancy), and establish relationships~(\egs is-similar-to, refines)
between these patterns.
The patterns they discuss are listed in \Cref{tab:rq5-cat-dp}.
A component model describes the design underlying each pattern.
Graphs represent the safety tactics.  GSN diagrams convey arguments
that a specific pattern implements a specific safety tactic and,
therefore, meets the top-level claim ``the system maintains its safety
functionality'' in an appropriate context.

\paragraph{Arguments}

\citet{Hawkins2009} explore a way to establish confidence for software
safety arguments.  They observe that certainty about claims made by a
software safety argument cannot be reached.  However, sufficient
confidence about these claims is required and assurance deficits
arising from uncertainties have to be made explicit~(\cf
\citep{Kelly1997}).  The authors apply deviation-style analysis~(\ies
HazOp) to identify assurance deficits and determine importance and
impact of each deficit.  They present five GSN
patterns~(\cf\Cref{tab:rq5-cat-ap}).

\citet{Palin2010} describe how safety cases can justify automotive
safety using a vehicle safety argument and 12 low-level
patterns~(\cf\Cref{tab:rq5-cat-ap}).  They combine their patterns
using ``supported by'' and ``in context of'' links.  They apply their
approach to an automotive start/stop system instantiating the
\emph{risk management argument} and the \emph{risk mitigation
  argument} patterns.  Their pattern catalog aims at the reuse of
arguments and the integration of design and safety activities.

\subsection{\RQRef{7}: Which technologies are abstracted by the patterns?}
\label{sec:rq-7-abstraction}

To understand the abstraction of the patterns and their applicability,
we classify them into three technology domains: \emph{software},
\emph{electrical and electronic hardware}, and \emph{mechanical
  hardware}.  Software includes programs and data structures.
Electrical and electronic hardware includes electrical, electronic,
and programmable electronic components~(\egs micro-processors, network
and communication hardware, field-programmable gate arrays).
Mechanical hardware includes, \eg metal frameworks, hydraulics, and
gear technology.

Models of safety concepts that abstract from several technology
domains and coherently integrate these domains are useful for the
evaluation of safety \wrt the system
perimeter~(\Cref{sec:from-designs-spec}).  The range of technologies a
model of a safety concept covers corresponds to the abstraction
available to reason about safety.  In summary, multi-domain
abstractions make it easier to reason about safety as a behavioral
constraint~\citep{Leveson2012}.

Several studies discuss their approaches by covering \emph{all three
  technology domains} \citep{Dardar2012, Denney2013a, Habli2010a,
  Kehren2004, Mahemoff2001, Mueller2012, Palin2010, Rauhamaeki2012,
  Rauhamaeki2015, Tan2009, Gleirscher2017}.

The following investigations cover the classical technology domains
considered in embedded systems engineering, \emph{software and
  electrical and electronic hardware}: \citep{Armoush2010,
  Baleani2003, Delange2009, Domis2009, Kelly1998, Miller2009,
  Pont2004, Preschern2013a, Preschern2013b, Preschern2013c,
  Preschern2014, Sun2010a, Tan2015, Trindade2014, Wu2003, Knight2012a,
  Littlewood2012, Delmas2017}.

\subsection{\RQRef{6.1}:  Which tactics are incorporated by the
  patterns?}
\label{sec:rq-6.1-principle-dep}

\begin{table}
  \caption{\RQRef{6.1}: Selections of studies by general principle or tactic
    \label{tab:rq6-1-selections}}
  \footnotesize
  \begin{tabularx}{1.0\linewidth}{L{.2}L{.8}}
    \toprule
    \textbf{Principle}
    & \textbf{Selection of Studies}
    \\\midrule
    Fault detection / condition monitoring
    & \cite{Crenshaw2006, Delange2009, Liu2008, Palin2010,
      Saridakis2002, Trindade2014, Littlewood2012}  
    \\
    Fault avoidance / simplicity and substitution
    & \cite{Delange2009, Pont2004, Rauhamaeki2012, Sun2010a, Tan2015,
      Grunske2003, Mahemoff2001, Basir2010} 
    \\
    Fault containment / masking, redundancy, recovery
    & \cite{Armoush2008, Iliasov2008, Saridakis2002, Littlewood2012,
      Chen2016, Mueller2012, Palin2010, Trindade2014, Baleani2003,
      Orlic2007, Rauhamaeki2015, Mahemoff2001}
    \\\midrule
    Fail-safe
    & \cite{Baleani2003, Bozzano2013, Eloranta2010, Gobbo2001,
      Palin2010, Preschern2013a, Preschern2013b, Preschern2013c,
      Preschern2014, Rauhamaeki2012, Saridakis2002, Knight2012a,
      Littlewood2012}
    \\
    Behavioral constraint
    & \cite{Crenshaw2006, Mueller2012, Rauhamaeki2012, Sun2010a, Trindade2014} 
    \\
    Preventive safety
    & \cite{Crenshaw2006, Mueller2012, Palin2010, Tan2009}
    \\
    Passive safety
    & \cite{Flammini2014, Palin2010}
    \\
    Stabilization
    & \cite{Crenshaw2006, Eloranta2010, Jain2012, Mueller2012,
      Rauhamaeki2012} 
    \\\bottomrule
  \end{tabularx}
\end{table}

We classified the patterns according to tactic taxonomies.  These
taxonomies help relating dependability principles as recommended by
standards, \eg fault-tolerance measures in IEC~61508.  Motivated by
such standards, \citet{Wu2003} elaborated a \emph{design tactic
  taxonomy} later refined by \citet{Preschern2013b}.
\Cref{fig:tactictaxonomy} arranges these and further tactics into an
extended design tactics taxonomy.  Additionally, the \emph{taxonomy of
  argument patterns} by \citet{Luo2016} distinguishes product-based,
process-based, compliance, and confidence arguments.

The following works include a comparison of fault avoidance,
detection, and containment principles: \cite{Armoush2010, Knight2012a,
  Liu2008, Mahemoff2001, Preschern2013b, Basir2010, Wu2003,
  Preschern2014, Larrucea2016, Orlic2007, Preschern2013c, Palin2010,
  Preschern2013a}.  These studies helped us to validate the
relationships in \Cref{fig:tactictaxonomy}.
\Cref{tab:rq6-1-selections} points to studies for the more general
principles (framed in boxes in \Cref{fig:tactictaxonomy}).

\subsection{\RQRef{12}: How are the patterns and tactics related to
  each other?}
\label{sec:rq12-patternrelationships}

\begin{figure}
  \includegraphics[width=\linewidth]{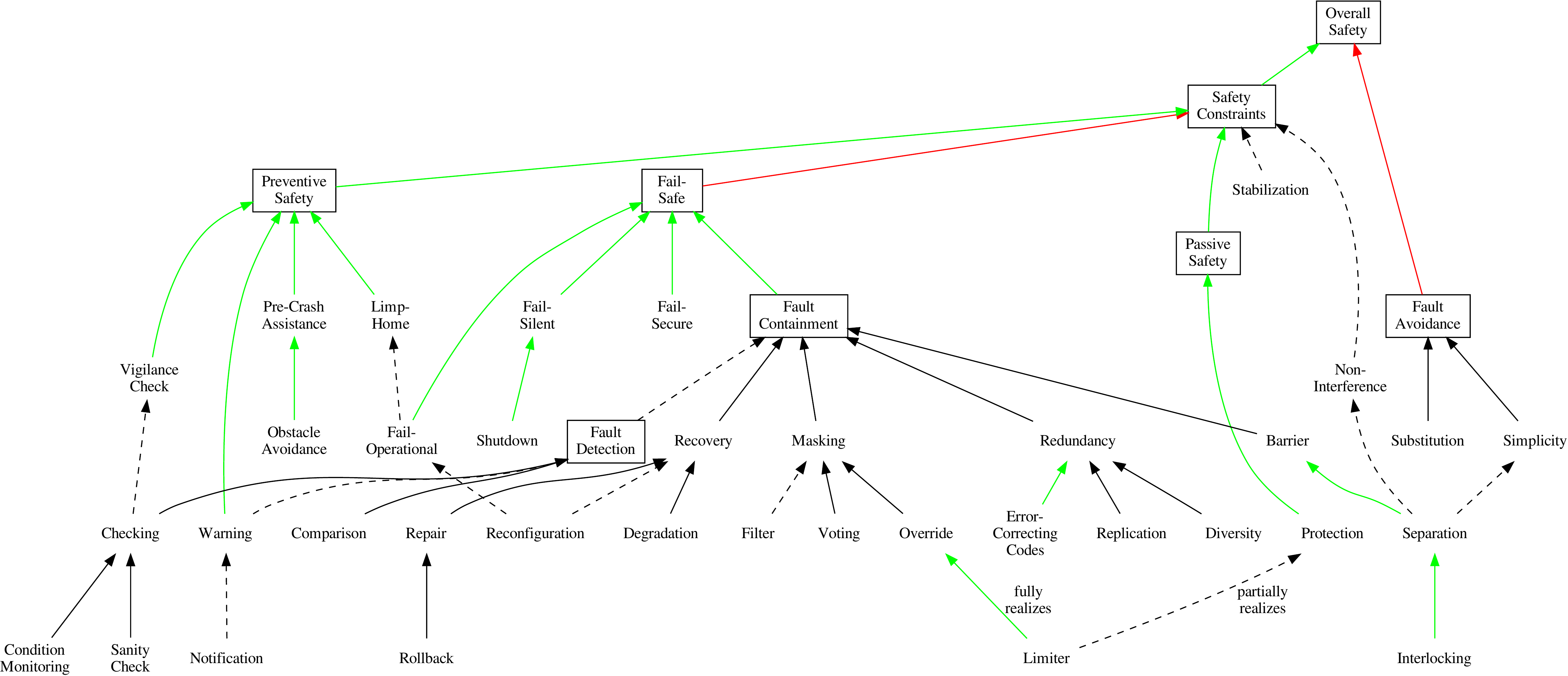}
  \caption{Overview of specification and design principles, comprising behavioral
    and decomposition tactics
    \label{fig:tactictaxonomy}}
\end{figure}

\Cref{fig:tactictaxonomy} depicts important relationships that
generally hold between the tactics:
\begin{itemize}
\item Solid arcs indicate that one tactic \emph{fully
    realizes} another tactic and dashed arcs that one tactic
  \emph{partially realizes} another tactic.  
\item Solid black arcs denote the \emph{generalizes} relationship
  between the tactics inherited from \citet{Preschern2013b,Wu2003}.
\item Solid green arcs signify relationships identified as a
  result of the survey.
\item Dashed black arcs indicate relationships that either include
  general principles or express indicate multiple inheritance.
\item Solid red arcs describe relationships that we believe are
  important but remain unclear or less clear from the surveyed
  studies.
\item Tactics only connected with dashed arcs indicate generic
  principles independent of safety.
\end{itemize}
\Cref{fig:tactictaxonomy} does not claim to present a complete set or
orthogonal basis of tactics and relationships.

\paragraph{Decompositions, Behaviors, and Specifications}

\Cref{tab:rq5-cat-dp} lists studies providing an overview of designs
used in safety engineering.  The column ``Fail-safe/fail-over''
indicates that designs are related to general behaviors through the
tactics they incorporate.
However, missing from many studies is a model (\egs a state machine)
describing the impact of the corresponding designs on system-level
properties (\egs a behavioral constraint in form of a temporal logic
formula).

If one of the designs in \Cref{tab:rq5-cat-dp} is chosen to enhance a
(component of a) system towards safety then, during assurance, we need
to be able to answer how the safety of this system is impacted by this
design.  Given the terminology in \Cref{sec:engineeringsteps}, using
such a design requires several claims to be substantiated, \eg that
the design improves the reliability of the system's components, that
the design improves the reliability of the whole system, or that the
design improves the safety of the whole system.  If these patterns can
be represented by behaviors then they can be verified against the
safety specification.

As summarized in \Cref{fig:tactictaxonomy} and accommodating
\citeauthor{Leveson2012}'s framework, behavioral constraints represent
the most general tactic to represent safe states of systems in their
operational environment.  This principle can be decomposed according
to \Cref{fig:tactictaxonomy}, \eg into fault detection (\egs condition
monitoring, limiters) and containment tactics (\egs recovery).

\paragraph{Decompositions and Arguments}

The \emph{fail-safe principle}~(\Cref{sec:safetypatterns}) incoporates
the ability of a system to maintain or achieve a safe state in case of
failure, \eg by shutting down the system or certain functions of it
(fail-silent) or by reconfiguring a systems' internal operational
state (fail-operational).  This principle can, \eg be realized by
redundancy, recovery, masking, and barrier.
From the studies mentioned in \Cref{sec:rq-6.1-principle-dep}, we know
that \emph{condition monitoring} is usually combined with
\emph{redundancy}, \emph{recovery}, and \emph{masking} to design
effective safety concepts.

\Cref{tab:rq5-cat-ap} provides examples of argument patterns~(\egs
decomposition arguments, argument modules), classifies them according
to their abstraction, their type~\citep{Luo2016}, and whether their
top-level claim contains a safety constraint (\egs a contract).
Furthermore, the table indicates relationships to design tactics~(\egs
passive and preventive safety).

\paragraph{Specifications and Arguments}

Specification patterns support the specification of behavioral
constraints, particularly, safety contracts.  Design patterns deal
with the behavioral and structural decomposition according to specific
decomposition criteria (\egs separation).  Of interest to overall
safety---as indicated in \Cref{fig:overview}---is to establish the
argument that the safety-enhanced design fulfills the safety
specification.
\Eg\citet{Sljivo2017} show, based on safety contracts, how
specifications and assurance arguments can be related and composed.

\subsection{\RQRef{6.2}: Which behaviors are guaranteed by the
  patterns?}
\label{sec:rq-6.2-principle-beh}

The analysis in \Cref{fig:tactictaxonomy} results in behavioral
constraints usually being a combination of \emph{fail-safety}~(both, fail-silent
and fail-operational), \emph{prevention}~(\egs by obstacle avoidance
or emergency braking), \emph{protection}~(\egs by limiters or
airbags), \emph{non-interference}~(\egs by interlocking), and
\emph{stabilization}~(\egs stability control of vehicle dynamics).
The \emph{prevention}~(particularly, vigilance check and obstacle
avoidance) and \emph{limiter} tactics usually cover mechanical
hardware with their abstractions.

The safety specification may more or less directly refer to the risk
class to be handled, \eg behavioral hazards, undesired interference,
and technical
defects~(\cf\Cref{sec:rq-4.1-mitigation}). %
If a safety-enhanced decomposition integrates one of the patterns, we
expect the behavior of this decomposition to fulfill the safety
specification.

Formal methods have proven to be very useful for the mathematical
study of whether a design fulfills a safety specification, \ie a
specific behavioral constraint.  Formal models of patterns along this
paradigm are discussed by \citet{Antonino2015,Basir2010,Ebnenasir2007,
  Jackson2001,Jain2012,Mueller2012,Riera2014,
  Sljivo2017,Sun2010a,Trindade2014}.

\subsection{\RQRef{8}: How is security addressed in the studies?} 
\label{sec:rq-8-secsafe}

Several of the surveyed works discuss security as a safety-critical
system property, particularly, undesired interactions between safety and
security, hazardous influences of security on safety or vice
versa.

\paragraph{Property Specification}
\label{sec:security-spec}

Relating security to other system properties,
\citet[Ch.~2.7.4]{Knight2012a} observes that ``security is inherently
a composite'' property whereas safety is discussed as a special aspect
of dependability dealing with technical defects with severe
consequences.
With their pattern, \citet{Kajtazovic2014} suggest that
contracts as a specification style can be used to specify both safety
and security properties.  \citet{Radermacher2013} propose a meta-model
facilitating the assignment of security and safety properties to
decomposition patterns.  Mixed-criticality encompasses interactions of
safety and security, such as mentioned by \citet{Nicolas2017}.
\citet{Asnar2011} present a modeling and specification framework for
security requirements that have to be verified for safety-critical
distributed information systems. %

\paragraph{Computing Architecture and Technology}
\label{sec:security-arch}

\citet{Zalewski2001} explains possibilities of intrusion into a
control computer.
\citet{Althammer2008a} describe a distributed multi-level security
architecture and summarize how their approach realizes modular safety
case construction including security arguments.
\citet{Wu2011} indicate how separation properties of a safety kernel
are derived from information system security mechanisms.
\citet{Littlewood2012} justify that the monitoring of 1-out-of-2
designs can indirectly mitigate consequences of certain security
attacks.
\citet{Larrucea2015a} discuss secure memory access in a
commercial-off-the-shelf processor. %
Additionally, \citet{Larrucea2016} support the combined handling of
safety, security, and real-time aspects through separation mechanisms
of a network-on-chip design pattern.
\citet{Cimatti2015} describe a fail-safe concept called
\emph{fail-secure} whose task is to bring a failed system to a state
without enabling security breaches.  The authors demonstrate how their
pattern can be automatically proven against safety/security contracts.
\citet{Nasser2017} exploit security attacks---\eg denial-of-service
and resource exhaustion---to introduce critical faults forcing a
system into its \emph{safe state}. \citeauthor{Nasser2017} observe that
safety mechanisms are perfect attack surfaces for such attacks.

Moreover, investigations by \citet{Aven2007,Novak2010,Eames1999}
suggest that \emph{undesired interactions} can occur in both
directions, \ies security attacks can not only create safety hazards
but safety measures can also result in security vulnerabilities.
Because of the increasing complexity and connectivity of control
software, such interactions could be reduced by integrated
safety/security concepts.

\paragraph{Engineering Process and Transformations}
\label{sec:security-process}

\citet{Hill2008} consider security as a part of the ``Product
Engineering Class'' of their ``software safety risk taxonomy.''
\citet{Kreiner2015} explains an architecture management procedure
applicable to handle both safety and security.
\citet{Preschern2014} compare several safety development methods
highlighting how three of them support a combined view of safety and
security: the ``Safe Control Systems'' method~\citep{Hauge2013}, a
method for the development of trusted applications for resource
constrained embedded systems \citep{Hamid2013-ModelDrivenEngineering},
and a method by the authors themselves~\citep{Preschern2013a} as
described later.
\citet{Castellanos2013} present an approach to the transformation of a
given safety-critical architecture into a security-enhanced
architecture.  They demonstrate their approach by transforming an
architecture model %
such that separation properties are fulfilled.

\paragraph{Argumentation}
\label{sec:security-arg}

\citet{Preschern2013a} extend their catalog of decomposition
patterns~(\Cref{tab:rq5-cat-dp}, \citet{Preschern2013c}) by
product-based \emph{security arguments} that disclose how typical
vulnerabilities could threaten implementations of these
decompositions.  These arguments refine a generic argument structured
according to a security analysis of the decomposition.  For
demonstration, they apply the STRIDE\footnote{The STRIDE approach
  encompasses the techniques of Spoofing, Tampering, Repudiation,
  Information disclosure, Denial of services, and Elevation of
  privilege.} analysis~\citep{Shostack2014} to identify threats using
data flow diagrams and add an argument against all identified threats.
The authors explain their approach using a substation automation
device case study from the railway domain.  This work integrates
safety and security by taking into account, at a pattern level, how
security threats could negatively influence safety properties.

\citet{Netkachova2015} describe an approach to identify interactions
between security and safety and to resolve conflicts leading to a
security-informed safety case.
\citet{Larrucea2017} propose a modular argument for a 
mixed-criticality design pattern.

\subsection{\RQRef{2}: Which models are used to describe the patterns?}
\label{sec:rq-2-language}
\label{sec:rq12-1-langbycat}

\begin{table}
  \caption{\RQRef{2}: Selection of studies by modeling 
    paradigm, formalism, and notation 
    \label{tab:rq2-selections}}
  \footnotesize
  \begin{tabularx}{1.0\linewidth}{L{.2}L{.8}}
    \toprule
    \textbf{Modeling Paradigm}
    & \textbf{Selection of Studies}
    \\\midrule
    Relational (\egs B, Z)
    & \cite{Gobbo2001, Sun2010a,
      Sun2014, Delmas2017} \\
    Propositional (\egs temporal logic)
    & \cite{Orlic2007, Steiner2011a, Trindade2014, Dias2011,
      Kehren2004, Basir2010, Sun2010a, Tan2015, Miller2009, Petroulakis2016} \\ 
    HOL (\egs PVS, Maude)
    & \cite{Steiner2011a, Dias2011, Sun2010a} \\
    Transition system, event structure (\egs AltaRica, Event-B)
    & \cite{Ball2009, Pereverzeva2012, Kabir2015, Iliasov2008,
      Kehren2004, Lin2015, Miller2009, Orlic2007, 
      Tan2015, Lopez-Jaquero2012, Randell1975, Nasser2017} \\
    Petri net
    & \cite{Belli1991, Flammini2014} \\
    Probabilistic (\egs Bayesian network)
    & \cite{Armoush2008, Armoush2010, Littlewood2012, Chen2016,
      Delmas2017, Bateman2006,Denney2012,Kramer2012, Peng2013,
      Zeng2016} \\ 
    Differential equation
    & \cite{Mueller2012} \\
    Structure, flow decomposition (\egs fault, signal, data, material)  %
    & \cite{Nasser2017,Alho2011,Baleani2003,Delmas2015,Grunske2003,
      Liu2008, Mahemoff2001, Natarajan2000, Pont2004, Preschern2013b,
      Rauhamaeki2012, Rauhamaeki2015, Tan2009, Luo2017, Bozzano2013, Preschern2014,
      Domis2009, Armoush2008, Armoush2010, Knight2012a} \\
    Pattern meta-model, tactic taxonomy
    & \cite{Khalil2014b, Preschern2013b, Radermacher2013,
      Vepsaelaeinen2014, Wu2003, Luo2016} 
    \\\midrule
    \textbf{Modeling Language} & 
    \\\midrule
    UML %
    & \cite{Fayad2003, Islam1996, Lopez-Jaquero2012,
      Miller2009, Saridakis2002, Giuntini2017} \\
    AADL, EAST-ADL
    & \cite{Delange2009, Cadoret2012, Castellanos2013, Oertel2014a,
      Penha2015a} \\ 
    ATL with OCL
    & \cite{Castellanos2013, Lin2015} \\
    Fault Trees
    & \cite{Antonino2015,Cimatti2015,Jaradat2015,Sljivo2017,Sorokos2016}\\
    GSN
    & \cite{Kelly1998, Kelly2006, Lin2015, Mayo2006, Netkachova2015,
      Palin2010, Preschern2013a, Basir2010, Preschern2013c, Weaver2003,
      Sljivo2017, Dardar2012, Denney2013, Denney2013a, Luo2016,
      Larrucea2016, Gleirscher2017, Sorokos2016, Habli2010a} 
    \\\bottomrule
  \end{tabularx}
\end{table}

\Cref{tab:rq2-selections} lists studies according to the modeling
paradigm (\egs relational, propositional, transition system) and the
modeling language (\egs UML, GSN) they use to present their approaches.

\subsection{\RQRef{15}: To what extent are known challenges covered by 
  the studies?}
\label{sec:rq15-challenges}

Inspired from discussions by
\citet{Cant2013,Langari2013,Laplante2007,Graydon2015,Graydon2017,Knauss2017},
we comment on some practical challenges we expect to be addressed by
research on safety concepts and assurance
cases. %

\subsubsection{Risk Identification and Classification}
\label{sec:risk-assessment}

\emph{Near-injectivity of Risk Classifiers}.
\citet[p.~4]{Cant2013} points to an issue with risk
classifiers~(\Cref{sec:safety}) also well-known from other
domains~\cite{Jarrett2010-Quantitativerisk} of risk analysis:
``different probability/severity pairs can be equated as having the
same level of risk.''\footnote{In a former practical course on
  applying hazard analysis techniques~\citep{Gleirscher2017}, our
  students raised this issue as well.}

We formally sketch this problem for the inclined reader: Let
$a,b\colon\mathbb{R}$ with $a < b$, $\mathit{Ev}$ the set of all
critical events, $e\in\mathit{Ev}$, $\mathit{Pr}(e)\colon[0,1]$ the
probability of occurrence of $e$, $\mathit{Sev}(e)\colon[a,b]$ a
severity of consequence measure for $e$, and a map
$\mathit{RL}\colon[0,1]\times[a,b]\to\mathbb{N}$.\footnote{We require
  $\mathit{RL}$ to range over a finite subset of $\mathbb{N}$, \ie to
  represent a finite partition of $[0,1]\times[a,b]$.}  Each natural
number can be assigned a set of directives for risk handling, \eg an
automotive \emph{safety integrity level}~(SIL) according to ISO~26262.
$\mathit{RL}$ encodes a domain-specific decision table that partitions
the combinations of $Pr$ and $Sev$ and maps $e$ to the risk level
according to this partitioning.  A rather simple way of risk
classification, \eg sometimes used in FMEA, is to multiply
$\mathit{Pr}$ with $\mathit{Sev}$ and map the result into an ordered
scale.

However, for $n\in\mathbb{N}$, the
inverse $\mathit{RL}^{-1}(n)$ can represent risk equivalence classes
\[
  \{(p,s)\mid
  p\in[0,1]\land
  s\in[a,b]\land
  \mathit{RL}(p,s) = n
  \}
\] being undesirably heterogeneous in their constitution, like pointed
out by \citet{Cant2013}.
For two fairly different risks of the same equivalence class, safety
analysts could make similar decisions about risk handling and safety
engineers could follow similar directives for risk handling.  Such
decisions and directives can involve, \eg the use of specific safety
patterns and specific parameters to instantiate these patterns.  In
the worst case, risks strongly differing in their $\mathit{Pr}$
and $\mathit{Sev}$ values are handled using the same design and
argument patterns where they should not.  The surveyed studies neither
address this issue nor do they provide information on how often it
occurs in safety practice.

\subsubsection{Risk Reduction and Assessment}
\label{sec:risk-reduction}

\emph{Architecture Hardening},
\emph{Integrity Level Decomposition}.
Hazard analysis drives the choice and design of safety concepts,
particularly, the enhancement of a given system with instances of such
concepts.  As described above, %
hazard analysis includes risk classification and the
assignment of SILs to \emph{critical items}, \ie the whole system or
its functions, channels, or components.  SILs represent safety
requirements to be fulfilled by such items.  Consequently, SIL
assignment corresponds to specification, and demonstrating that an
item achieves a certain SIL corresponds to assurance argumentation.
Standards such as ISO~61508 and ISO~26262 %
provide descriptions of SILs and, furthermore, suggest decomposition
schemes for SILs according to fault tolerance tactics (\egs
redundancy).
This shows how safety requirements decomposition is driven by certain
tactics~(\Cref{fig:tactictaxonomy}).  Usually defined for electronic
and mechanical hardware items, some standards (\egs IEC~61508) also
provide SIL classifications for software items.

In this context, risk reduction involves two main questions:
First, for architecture hardening, given an item $S$ known to have SIL
$x$, does the transformation of $S$ into $S_P$ using a decomposition
pattern $P$ lead to a level $x_P$ of $S_P$ strictly higher than $x$?
\Eg \citet{Grunske2003} and \citet{Delmas2017} investigate the idea
of transforming an architecture model by instantiating a safety
pattern and synthesizing a \emph{hardened} architecture.
Second, for integrity level decomposition, given an item $S$ assigned
SIL $x$ and decomposing $S$ into $S_1 \otimes_P S_2$ by applying a
decomposition pattern $P$, can the SILs of $S_1$ and $S_2$ be reduced
to levels $x_1$ and $x_2$ strictly lower than $x$?  This last question
leads to assumptions to be made about the context of $S$ and to
requirements to be verified of $S_1$ and $S_2$ and their composition
by $\otimes_P$.

\emph{Alignment of Risks with Safety Functions}.
System-level hazards are defined to have an impact on the system's
environment.  Whether or not such hazards are caused by component
failure,
studies about behavior patterns discuss what \citet{Cant2013}
describes as ``safety functionality in terms of the system
interface.''  He states that ``hazard analysis is inward-looking,
making it hard to describe safety functionality in terms of the system
interface.''  Certainly, hazard analysis should help with the
identification of safety requirements both at system and component
level and with decisions on the safety concepts to be used.  We
believe that hazard analysis based on behavior models can improve the
identification of safety requirements at the system interface and,
thus, improve the verification of the absence of system-level hazards.
\Eg\citet{Mahemoff2001} discuss patterns capturing ``safety-usability''
for interface designs and \citet{Murugesan2015} elaborate an extensible
state machine pattern suitable for identifying and reducing mode
confusions.

\emph{Certifiable Testing of High Automation}.
\citet{Knauss2017} empirically identify challenges in automated
vehicle testing, particularly,
\begin{inparaenum}[(i)]
\item the practical need of improved safety standards,
\item simple integrated models for deriving complete and sound test
  suites, and
\item the avoidance of re-certification.
\end{inparaenum}
We extend their investigation with 
the impact of safety design patterns on these challenges.  
\Eg product-based argument patterns
can be a complement to compliance arguments backed by procedural
standards~(i), design patterns
can simplify test suite decomposition through reuse~(ii) and,
moreover, reduce the fragility of safety certificates~(\egs by
argument modules and by using the separation tactic; iii).

\emph{Avoidance of Safety Antipatterns}.
\citet{Brown1998} discuss general antipatterns in software engineering
management and \citet{Laplante2007,Laplante2005} transfer these ideas
to various domains.  \Eg \citet{El-Attar2006} apply antipatterns to
software specification based on use cases and \citet{Moha2012} apply
antipatterns in architecture analysis.  However, we are unable to find
studies identifying ``worst practices'' in a safety-related 
context.

\subsubsection{Argumentation}
\label{sec:argumentation}

\emph{Objectivity of Risk Analysis}.
To reduce confirmation bias in an argument, \citet{Leveson2011}
proposes to \emph{negate the assurance claim}.  The integration of
hazard analysis techniques in the argumentation process can reduce
this problem, \eg fault tree analysis~(see, \egs\citet{Oertel2014a})
helps to identify how an undesired top-level event could possibly
occur and argue from~(reusable) countermeasures~(see,
\egs\citet{Gleirscher2017}).
The negated assurance claim %
would require an iterative argumentation process with risk reduction
after each iteration and a corresponding update of the argument.  This
update would consist in the weakening of the
claim %
and the pruning of the evidence for the mitigated hazard %
from the argument.  This process stops if no further evidence for
substantiating the negated claim can be found, the claim is ultimately
weakened, %
and all existing evidence has been removed.

\emph{Enrichment of Process-based Arguments by Product-based Arguments}.
\citet{Graydon2007} propose the on-the-fly construction of assurance
evidence during development.  Other studies discuss traceability (i)
between evidence and assurance claims by building safety arguments
from architecture models and component fault
trees~\citep{Domis2009,Sorokos2016} and (ii) between counter-evidence
and assurance claims by obstructing safety arguments through fault
injection and propagation and obtaining fault trees from computed
cut-sets~\citep{Oertel2014a}.
\citet{Basir2010} attaches meaning to assurance arguments by using
natural deduction and, this way, indicates how inconsistencies during
integration of evidence can be disclosed by theorem proving.

\emph{Validity of Assurance Assumptions and Models}.
\emph{Argument confidence}~(\egs completeness, soundness) is relying
on the validity of the models of the system under argumentation and
its environment and on the chosen abstraction.  \citet{Sun2011}
demonstrate how to justify the validity of hazard analysis results
to be employed in an argument.
In the maintenance of safety-critical electronic hardware (\ies
prognostics and health management), \citet{Zio2016} discusses
opportunities of using data from hardware health monitoring sensors
for the assessment and usage of predictive models.

\emph{Unambiguity, Soundness, and Completeness of Arguments}.
Formalization is known for the reduction of ambiguities, the support
of automation, and the evaluation of consistency of claims.  However,
it is still unclear whether \citep{Rushby2010} or not
\citep{Graydon2015}, and to what extent formalism in arguments can
increase their confidence.
In any case, the contract templates from \citet{Antonino2014} could be
formalized and assessed against completeness criteria.
\citet{Luo2017a} present a tool for confidence assessment of arguments
(based on degree-of-beliefs) without the necessity of having a formal
design model of the system.  Their approach seems useful in the
construction of abstract arguments to be re-assessed once such models
are available.

\emph{Readability, Reduction of Complexity, and Maintainability of Arguments}.
\Eg
\citet{Despotou2008} and \citet{Kelly2006} apply decomposition
criteria from architecture design to modularize assurance
arguments.
While GSN has been standardized and shown to be useful for the
representation of arguments~(\Cref{sec:rq-2-language}),
\citet{Denney2013a,Denney2016} and \citet{Matsuno2014} provide
formalizations of the GSN syntax to support composable GSN-based
patterns and to support automated consistency checking for argument
construction and maintenance.
We believe, their decomposition and consistency checking could make
use of behavior models of the system under argumentation.  This was
shown in goal-oriented and model-based requirements
engineering~(\egs\cite{Lamsweerde2009}) %
and in contract-based software and systems
engineering~(\egs\citep{Meyer1992,DBLP:series/natosec/Broy11}).
Studies such as, \eg\citet{Armoush2010,Orlic2007,Sun2010a,Basir2010}
also discuss this direction.  Overall, correctness proofs based on
behavior models could help justify the decomposition of the
corresponding GSN arguments.  Using behavior models with formal
contracts would allow arguments to scale up to the assurance of large
systems.

\section{Discussion}
\label{sec:discussion}
\label{sec:findings}
\label{sec:recomm-based-our}

The \Crefrange{sec:find-from-syst}{sec:find-regard-empir} present our
findings and suggestions for future research based on the survey
results.  In \Cref{sec:limitations-survey}, we
discuss validity threats from literature search and systematic mapping
with a potential impact on the quality of this survey and describe our
measures to minimize these threats.  \Cref{sec:exper-with-surv}
summarizes our experiences with the survey method. 

\subsection{Findings and Suggestions from Systematic Mapping}
\label{sec:find-from-syst}

The applications discussed in the studies cover many areas of
safety-critical embedded control systems. %
Patterns are presented at different stages during the system life
cycle ranging from early-stage requirements
engineering~(\egs\citep{Stalhane2016}) and architectural
design~(\egs\cite{Rupanov2012}) %
down to implementation and assurance.  Several studies capture at
least three of the four engineering steps we distinguish.
Of all categories, our survey targets design~(\ies
behavior, decomposition) and argument patterns.
Particularly, the surveyed decomposition patterns cover the whole
range of tactics from fault avoidance~(``correct by design'') and
fault detection to fault containment, however, with a clear focus on
checking, recovery, and redundancy.

Notations like UML, GSN, and generic component diagrams appear most
frequently in the presentations of the concepts.  GSN dominates the
studies and is discussed as an intuitive way of visualizing arguments.
Formalism is rarely used in the studies to assess the concepts.

Overall, we found interesting research towards a solution of each of
the listed challenges.  However, risk
classification, %
integrity level decomposition, %
and the identification of antipatterns %
were discussed the least among the surveyed studies.  Only few of the
works present validations and practical evaluations based on empirical
and formal results.

The surveyed works cover many of the internationally relevant venues
for design patterns and for dependability engineering%
.

We strongly encourage the use of a structured abstract and a standard
document template.  Document templates used in the studies helped us
to evaluate the described patterns and relate them to each other.
Templates can also be a first step towards formalization.

\subsection{Towards a Unified Pattern System}
\label{sec:unification}
\label{sec:find-regard-patt}

Relationships between design patterns have, \eg been elaborated by
\citet{Preschern2013b}.  From the studies, we identified further
relationships between design and argument patterns as shown in the
\Cref{tab:rq5-cat-dp,tab:rq5-cat-ap}. %

Several patterns facilitate the abstraction from software, hardware,
and mechanical aspects.  However, the level of abstraction used in the
studies varies strongly.  Moreover, the studies suggest that there are
several related notions of safety.  Formal models would allow the
comparison of these notions.
Without formalization, relationships (\egs similarity, equivalence,
refinement) between the patterns as well as the meaning of their
composition remains unclear.

\Eg an integration of the approaches of
\citet{Rauhamaeki2012} and \citet{Sljivo2017} would require unified
notions of a safety function, a safe state, and relationships between
the presented patterns.
Beyond the identified relationships, refinements between
design-time~(\egs hardware platform substitution) and run-time~(\egs
multi-channel redundancy with voting) fault
prevention~\citep{Grunske2003} could help bridging important reasoning
steps in assurance arguments.
The many country-specific \emph{signaling/interlocking practices} in
the railway domain further motivate unification.  Because railway
regulation is also accident driven~\citep{Havarneanu2015}, relating
these practices and deriving a hierarchy of interlocking patterns
could even lead to an exchange of lessons learned~(from train
accidents) between those countries.

To improve the comparability and unification of patterns, they could
be modeled in a \emph{common formal framework}, %
\eg based on relations~\citep{Stepney2003} or temporal
logic~\citep{Dwyer1999}.
Such a formal basis could help clarifying and establishing new pattern
relationships, such as \emph{inheritance}, to elaborate pattern
variants, and establish proof hierarchies (\egs envisaged
in~\citet{Marmsoler2017a}).

\subsection{From Tactic Taxonomies to Proof Systems}
\label{sec:formalization}

Tactic taxonomies can be seen as a first step towards a theory of
safety concepts.  Such a theory could bridge the gap between
decomposition tactics~\cite{Wu2003,Preschern2013b} and argumentation
tactics~\cite{Luo2016} and comprise a framework for the construction
of verifiable safety arguments.  According to
\citeauthor{Rushby1994}'s classification of critical
properties~\citep{Rushby1994,Rushby2010}, a taxonomy can foster reuse
in the verification of how safety design patterns contribute to
safety.  A taxonomy can, moreover, identify the role these patterns
play in high-confidence argumentation.

Many of the decomposition patterns are also known as reliability
patterns. %
Data in \citet{Gleirscher2018-safetysurvey} suggests that
practitioners are aware of the subtle differences between safety and
reliability.  This distinction has not been rigorously clarified in
the studies, including the pattern
catalogs~\citep{Grunske2003,Preschern2013c}.  However, in
\Cref{sec:betw-spec-behav}, we distinguish between safety as a strong
property and reliability as a weaker, more general property.

Safety, reliability, and security can be formulated as behavioral
properties of a system, as explained in
\Cref{sec:background-behavior}.  Hence, a framework based on
\Cref{fig:tactictaxonomy} would be helpful to verify how
safety-enhanced designs fulfill a given set of behavioral constraints,
\ie a safety specification~(\Cref{fig:overview}).

Inspired from discussions by \citet{DBLP:series/natosec/Broy11} and
\citet{Rushby2011},
verification results~(\egs safety integrity level or, more generally,
property decomposition and preservation) could be lifted to the
pattern level and, this way, form a pattern-based proof
system. %

\subsection{Argumentation by Contract Verification}
\label{sec:formalization}

The view of safety as a behavioral
property~(\Cref{sec:viewsofsafety}) not necessarily related to
system failure is rarely discussed in the studies. %
In the light of safety as a behavioral property~\citep{Leveson2012},
many of the design patterns we surveyed are missing a 
corresponding \emph{contract}~(\Cref{sec:safetypatterns}) that allows to
reason~\citep{Rushby2010,Rushby2011}
about how a design impacts safety and reduces hazards.  Instead,
safety often seems to stay implicit in a pattern application.

For the construction of high-confidence arguments, the use of behavior
models could help connect designs and arguments.  Moreover, formal
models could be used to clarify the semantics of certain parts of
arguments.
\Eg the work by \citet{Basir2010} shows a fruitful connection
between %
theorem proving and argument construction.
Moreover, we believe that the definition of contracts for patterns is
necessary for successful reuse and the construction of high-confidence
arguments.
\Eg for safety mechanisms in automotive engineering, mechatronics, and
robotics (\egs anti-lock braking systems, vigilance checks
\cite{Lin2011,Hirata2014,Utsumi2013}), it can be useful
to have property specifications (\egs contracts) as assurance
claims. %

A fail-safe system is required to not create a hazard on the
occurrence of covered failures.  What does that mean? What is the
considered system perimeter?  Can we deduce from a model that failure
consequences have actually been covered beyond the requirements? Can
we deduce safety of a system in the failed state from this model?  We
could, \eg do so using a contract that specifies the weakest context.
But how much do we need to know about this context to specify safety
requirements in form of safety contracts?

Many of the design patterns based on the fail-safe and
fail-over tactics (\Cref{tab:rq5-cat-dp}) model decompositions that
can be characterized by a corresponding behavior.  In fact, some
designs combine decomposition and behavior, \eg the ``distributed
multiple independent levels of security'' design is formally modeled
and verified against a contract as demonstrated in
\citep{Cimatti2015}.
Moreover, \citet{Dong2003} show the possibility of design patterns to
be associated with contracts inducing expressions in
\citeauthor{Milner1995}'s Calculus of Communicating
Systems~\cite{Milner1995}.

Furthermore,  \emph{contracts with violation handling} (\ies
specifications with assumptions on the environment) could allow
critical contract violations to be handled by weakening contracts
and instantiating corresponding pairs of patterns and contracts.
\begin{quotation}
  This way, the instantiation of a safety pattern for a system can
  address a violation of the assumption of a contract (\egs a failure mode of
  the environment) or a violation of the guarantee of a contract (\egs
  a failure mode of the system).
\end{quotation}
Contracts with violation handling as, \eg formalized by
\citet{DBLP:series/natosec/Broy11}, can get highly important
when system (of systems) complexity boosts latent systematic
defects. %

\subsection{Hazard Analysis and Risk Assessment}
\label{sec:find-regard-patt}

It is important to keep track of how low-level hazard analysis impacts
design decisions for the overall system and how system-level hazard
analysis drives low-level design decisions.
Because risk priorities %
correspond to integrity levels %
and integrity levels motivate the choice of decompositions, the issue
of homogeneous risk classifiers is worthy to be further explored at a
pattern level.  Injective risk classifiers of the type
$RL\colon Ev \rightarrow \mathbb{N}$ could aid in more distinctive
risk handling by providing information about the type of risk beyond
just $Pr$ and $Sev$ for classification.

Related to risk classification is the treatment of critical properties
of different nature, \eg security and safety, usability and safety, or
usability and security.  Such properties are usually allocated to
various overlapping fragments of a system.  Particularly, if
interaction cannot be handled by conservative forms of physical
separation, the avoidance or handling of \emph{undesired interactions}
between these system fragments forms a safety requirement.

Finally, formalism could also provide evaluation criteria for argument
patterns, \eg to identify confirmation bias. %
For learning from failure, a
stronger integration of incident and accident
research~\citep{Havarneanu2015} as recommended by \citet{Leveson2012}
seems necessary to report on \emph{antipatterns} and to improve
existing safety concepts.

\subsection{Tools Support for Pattern Systems}
\label{sec:find-regard-modell}

Verification frameworks could benefit from process and tool support
based on patterns and support safety engineering as an integral part
of the system life cycle.
The integration with widely known system models and
formalisms~(particularly, to address
cross-disciplinarity) %
could improve automated construction, semantics-based checking and
verification, and maintenance of assurance results.  Such approaches
would offer tool support for analyzing
risk, %
determining safety integrity
levels %
and, for architecture hardening at
scale. %

The relationship between assurance arguments~\citep{Palin2010} and
requirements traceability is discussed
by~\citet{Antonino2015,Sljivo2017}.
Beyond traceability models, support for automated traceability is
desirable in the construction and maintenance of arguments from
evidence~(\egs implementation and verification of countermeasures)
towards claims~(\egs system safety requirements).

\subsection{Empirical Assessment and Standardization of Pattern Use}
\label{sec:find-regard-empir}

Although safety patterns represent ``best practices'' or
``proven-in-use'' concepts, we do not know whether the
discussed patterns fulfill the predicates ``best'' or ``proven.''  The
examples, case studies, experience reports, and discussions studied
help with understanding the approaches and suggest these predicates
hold of decomposition patterns more than of argument
patterns. %

The standard IEC~61508 (part~4) defines \emph{proven in use} for an
element as the demonstration
\begin{quote}
  ``... that the
  likelihood of dangerous systematic faults is low enough so that
  every safety function that uses the element achieves its required
  safety integrity level.''
\end{quote}
\citet{Ehrenberger2016} and \citet{Schaebe2015} discuss the
application of stochastic models (here, \textsc{Poisson} processes as
proposed by Littlewood~\cite{Littlewood1991}) in the transfer of
reliability characteristics between different contexts.  Consistently,
the standards ISO~26262, IEC~61508, and ISO~61511 refer to techniques
such as ``field monitoring'' to collect statistical evidence for
arguments of items to be reused in certification.

For the transfer of the ``proven in use'' concept to the field of
argument patterns, it can be helpful to evaluate how and where the
surveyed argument patterns have been practiced and declared as fit for
reuse in certification.
For this, a database using the structure of
\citeauthor{Kajtazovic2014}'s approach could be used to register
``system elements out of context'' (\cf ISO~26262) or other reusable
system components.  For each item in the database, the contract could
store safety properties and a decomposition.

Furthermore, the inclusion of formal safety patterns for
specification, design, and assurance in industrial standards could
guide the representation of all safety stakeholders' expectations on
argument confidence. %
\Eg the design patterns to cover random faults in electronic hardware
components in IEC~61508 Parts~2 and~7 (diagnostic techniques and
measures) could be extended by corresponding argument and
specification patterns.

The relatively low number of validation, evaluation, and experience
studies hampers successful knowledge transfer between industry and
academia.  Although the studies convey interesting results, some
studies are, thus, difficult to replicate and transfer.
Hence, we stress the necessity to collect evidence on pattern
applications from further application domains~(\egs robotics and
intelligent autonomous systems) and in form of validation and
evaluation works, preferably, controlled experiments using formal
methods.

GSN-based assurance cases~\citep{Hawkins2009} have received a lot of
attention by researchers leading to a body of knowledge ready to be
empirically assessed, further integrated with research from other
fields (\egs formal methods, software engineering, reliability
engineering), and further transferred into assurance practice
(\egs through stronger inclusion in standards).
These observations confirm and generalize
\citeauthor{Graydon2015}'s~\citep{Graydon2015} question of sufficient
evidence on the quality, properties, and practical effectiveness of
formal assurance arguments.

\subsection{Limitations of the Survey}
\label{sec:limitations-survey}
\label{sec:threats}

Our search was focused on patterns, concepts, measures, and mechanisms
in the design and assurance of safety-critical systems%
.  The ambivalent meanings of the words ``pattern'' and ``safety''
lead to many search results across many domains and disciplines.
Differences in the search engines first required us to relax and then
constrain the search expressions.  \Eg we pruned the results by
excluding the terms ``food, medic, bio, service-oriented architecture,
SOA, web.''
Moreover, we exclude gray literature and non-peer reviewed literature%
, but include peer-reviewed and archived literature from the same
authors if available.  However, we reduced the risk of missing
relevant work and outdated search results by snowballing%
\footnote{Manually picking studies from bibliographies of selected key
  studies.} and an update
of our search%
.
We eliminate inaccuracies during data extraction, coding, and
analysis by discussion, redundant extraction and coding, and
cross-checking data samples.

\subsection{Experiences with Systematic Mapping and Review}
\label{sec:exper-with-surv}

Systematic literature reviews~\citep{Kitchenham2007} are intended to
help with the comparison of similar studies~(\egs replication studies)
in a very specific research field and with identifying new research
directions in this field.

In our cross-disciplinary case, the effort of a systematic review to
compare heterogeneous studies and to identify interesting research
directions was too high to justify its benefit.  Our main
findings %
could have been identified more easily by relying on one or two
databases instead of four and by snowballing.  We thus confirm the
drawback of high costs as mentioned by \citet[p.~352]{Shull2008}.
Database search and filter turned out to be a weak instrument for the
control of the survey process to quickly converge in a complete
selection of coherent studies.  Search terms can play a minor role in
identifying relevant literature as the search can get unnecessarily
wide in case of many synonyms and biased in case of missing synonyms.

From four databases and from duplicate removal, we observe
that the gain in relevant studies diminishes with the addition
of a third and fourth database.  Our experience suggests that after an
initial search in one (or two) database(s), traditional manual
\emph{snowballing} %
seems to be most effective and can be supported by tools\footnote{See,
  \eg\url{https://openknowledgemaps.org/} or
  \url{https://www.semanticscholar.org}.} and by specific
search queries.  We have however not taken into account the simplifications
for systematic reviews recommended by \citet{Kuhrmann2017}.

The most difficult parts in exploring the literature of a
cross-disciplinary field is the identification of an interesting set
of survey questions.  Such questions typically arise from the review
process and often require changes of the search string because new
concepts and terms have to be taken into account.  This issue
necessitates search iteration and strengthens the case for snowballing.

\section{Conclusions}
\label{sec:conclusion}

In this survey, we summarize and evaluate research on reusable design
and argumentation concepts for the assurance of system safety.  We
present relationships between these concepts and derive suggestions
for future research.
Our work is based on a systematic map to reduce bias,
validate search criteria and databases, to track inclusion and exclusion
criteria, and to support reproducibility.

We identify a range of reusable concepts.  Most of these concepts can
be related to the overall aim of not violating safety constraints.
The difficulty of effective reuse of these concepts still sustains
with the problem that the various notions of system safety are not
well integrated.  The most relevant notion defines safety as a set of
behavioral constraints to be fulfilled by the system when operated in
its environment.  Additionally, safety of a system in operation depends on
the various properties of this system's components.  Hence, a reusable
safety concept needs to explain how it, when applied in a specific context,
provably contributes to the system's safety.  However, heterogeneous
formalisms hamper the comparison of the concepts across the surveyed
studies.

The construction of high-confidence assurance arguments requires
precise semantics of these concepts for the analysis of their
relationships and for the verification of their composition.  Safety
concepts could be modeled as \emph{pairs of design patterns and safety
  contracts} to facilitate compositional verification of their use,
\ie of their instantiation. Studies using formal~(\egs probabilistic)
models of these concepts are most promising to meet assurance proof
obligations of this kind.

To improve this situation, we identified three interesting research
directions:
First, the establishment of a body of knowledge unifying the concepts
shared between safety assurance, dependability and security
engineering, human factors engineering, and system accident research.
Second, the formulation of reusable concepts in a common semantic
framework.
Third, the transfer of this body of knowledge into engineering
practice to evaluate the effectiveness of these concepts and gain
feedback for research progress.

\paragraph{Future Work}
\label{sec:futurework}

Our survey could be extended for discussing safety and security.
Hence, a repetition of our study for \emph{security patterns} could
balance our safety perspective by a security perspective.  A similar
survey on ``security patterns'' could help to develop an integrated
view of security and safety~\citep{Steiner2013}.

An interesting direction for the development of an improved safety
pattern system could be a survey of \emph{safety-related
  antipatterns}.  This step might have to go along with empirical
studies of system safety practice and could take into account more
general work on design ``patterns'' ``templates'', ``models,''
``types'', ``forms'', ``tactics'', or ``styles''.
The survey of gray literature could take into account unpublished but
open material from industrial practice.

An extension of the presented tactic taxonomy could be
facilitated by additional studies on the reuse of designs and
arguments in, \eg robotics and intelligent autonomous systems,
public infrastructure automation, %
building automation in civil engineering, production automation, naval
systems, and control systems in construction machinery and utility
vehicles.

The classification in \citet[p.~208]{Pont2004} could help with a
refined analysis of the questions \RQRef{1}, \RQRef{5}, \RQRef{6.1},
and \RQRef{12} for software implementation patterns.  The authors
distinguish between patterns for software foundations, time-triggered
architecture for single/multi-processor systems, user-interface
components, serial-peripheral library, monitoring and control
components, and hardware foundations.

\begin{acks}
  This work is supported by the \grantsponsor{501100001659}{Deutsche
    Forschungsgemeinschaft
    (DFG)}{http://doi.org/10.13039/501100001659} under the
  Grants no. \grantnum{501100001659}{GL~915/1-1} and
  \grantnum{501100001659}{GL~915/1-2}.  Preliminary studies for this
  work were supported by the German automotive industry.  We would
  like to spend sincere gratitude to Ketil St{\o}len for helpful
  suggestions on restructuring our discussions and conclusions.
  Moreover, we are pleased to thank Jim Woodcock for several comments
  towards the completion of our analysis and the interpretation of our
  results.  Special thanks go to Maged Khalil for helpful feedback on
  improving the first stages of this survey.  Finally, we thank the
  anonymous reviewers for their guidance in preparing an improved
  version.
\end{acks}

\appendix
\bibliographystyle{ACM-Reference-Format} 
\bibliography{}

\section*{\appendixname}

\renewcommand{\appendixname}{}

\section{Nomenclature}
\label{sec:list-abbreviations}

See \Cref{tab:abbrev}.

\begin{table}[h]
  \centering
  \scriptsize
  \caption{Important abbreviations used in this article
    \label{tab:abbrev}}
  \begin{tabularx}{1.0\linewidth}{lX}
    \toprule
    AADL & Architecture Analysis and Design Language\\
    (A)SIL & (Automotive) Safety Integrity level \\
    ATL & Atlas Transformation Language\\
    EAST-ADL & EAST Architecture Description Language\\
    ETA & Event Tree Analysis \\
    FFA & Functional Failure Analysis \\
    FMEA & Failure Mode Effects Analysis \\ 
    FTA & Fault Tree Analysis \\
    GSN & Goal Structuring Notation \\
    HazOp & Hazard Operability (studies)\\
    HOL & Higher Order Logic \\
    IMA & Integrated Modular Avionics\\
    LTL & Linear Temporal Logic \\
    OCL & Object Constraint Language \\
    PVS & Prototype Verification System \\
    RE & Requirements Engineering \\
    SC & Safety Concept \\
    SE & Software Engineering\\
    STRIDE & Spoofing, Tampering, Repudiation, Information disclosure,
             Denial of service, Elevation of privilege \\
    UML & Unified Modeling Language \\
    \bottomrule
  \end{tabularx}
\end{table}

\end{document}